\documentclass[11pt,aps,preprintnumbers,nofootinbib,superscriptaddress,notitlepage,longbibliography]{revtex4-2}

\usepackage[T1]{fontenc}
\usepackage[english]{babel}
\usepackage{stix}
\usepackage{epsfig}
\usepackage{amsmath,bm,bbm,tensor,physics,dsfont}
\usepackage{graphicx}
\usepackage{verbatim}
\usepackage{csquotes}
\usepackage{simplewick}
\usepackage{appendix}
\usepackage[usenames,dvipsnames,svgnames]{xcolor}
\usepackage[normalem]{ulem}
\usepackage{hyperref}
\usepackage{cleveref}
\usepackage{orcidlink}

\definecolor{IAN}{RGB}{1,80,158}

\usepackage[framemethod=default]{mdframed}
\newmdenv[skipabove=7pt,
skipbelow=7pt,
rightline=false,
leftline=false,
topline=false,
bottomline=false,
backgroundcolor=gray!10,
linecolor=gray,
innerleftmargin=5pt,
innerrightmargin=5pt,
innertopmargin=5pt,
innerbottommargin=5pt,
leftmargin=0cm,
rightmargin=0cm,
linewidth=4pt]{eBox}

\newcommand{\md}{\mathrm{d}}

\begin{document}

\title{Ghost instabilities and strong coupling in quadratic non-metricity theories}

\author{\textsc{Alexander Ganz\,\orcidlink{0000-0001-7939-9058}}}
    \email{{alexander.ganz@itp.uni-hannover.de}}
    \affiliation{Institute for Theoretical Physics, Leibniz University Hannover, Appelstraße 2, 30167 Hannover, Germany.}

\author{\textsc{Marco Spinelli\,\orcidlink{0009-0008-4531-3728}}}
    \email{{marco.spinelli@iusspavia.it}}
    \affiliation{Department of Physics ``A. Pontremoli'', University of Milan, Via Celoria 16, 20133 Milano, Italy}
    \affiliation{IUSS Pavia, Palazzo del Broletto, Piazza della Vittoria 15, 27100 Pavia, Italy.}

\begin{abstract}
We revisit the framework of Newer General Relativity, defined by all independent quadratic invariants of the non-metricity tensor, including the unique quadratic parity-violating term. We analyze linear perturbations around a flat FLRW background and find that the theory generically exhibits ghost instabilities and/ or propagates more degrees of freedom than in the Minkowski limit, signalling strong coupling. There are two notable exceptions: the Symmetric Teleparallel Equivalent of General Relativity (STEGR) and the transverse-diffeomorphism-invariant gravity subclass, both of which are supplemented by the parity-violating operator. However, since the parity-violating term explicitly breaks (transverse) diffeomorphism invariance, we show, using both the Dirac-Bergmann procedure and the Cartan-Kuranishi algorithm, that the parity-violating extension of STEGR propagates eight degrees of freedom at the fully non-linear level. 
\end{abstract}

\maketitle

\section{Introduction}

Grounded in the equivalence principle, General Relativity (GR) establishes a profound connection between gravity and the geometry of spacetime, wherein the motion of matter is governed by the curvature of spacetime itself. In its standard formulation, spacetime is represented as a pseudo-Riemannian manifold equipped with the unique metric-compatible and torsion-free Levi-Civita connection. Subsequent developments have relaxed these geometric conditions on the affine connection to investigate the theoretical implications of introducing torsion and/or non-metricity. 

The earliest attempt to extend GR in this direction can be traced back to Élie Cartan, who, shortly after Einstein's original formulation, proposed that torsion might be associated with intrinsic angular momentum  \cite{Cartan1922,Cartan1923,Cartan1924,Cartan1925}. However, since the concept of electron spin had not yet been discovered at the time, Cartan’s idea was largely overlooked. In the 1950s, this line of thought was revived, leading to extensive investigations into the role of torsion and connection to formulating gravity as a gauge theory of the Poincaré group (see \cite{Hehl:1976kj} for a review). 

The concept of non-metricity has gained significant attention only in recent years, following the development of equivalent formulations of GR constructed either entirely in terms of torsion \cite{Ferraris:1982ez} or entirely in terms of non-metricity \cite{Nester:1998mp,BeltranJimenez:2017tkd,BeltranJimenez:2019esp} - nowadays commonly referred to as Teleparallel or Symmetric Teleparallel Equivalent of GR (TEGR or STEGR), respectively.  Note that, depending on the geometry of the connection, the prescription of matter couplings may become ambiguous and can lead to issues (see \cite{deAndrade:1997cj,BeltranJimenez:2020sih,So:2006pm,Delhom:2020hkb}  for a more detailed discussion).  

Recently, there has been a growing interest in developing modified theories of gravity based on these generalized geometrical frameworks and in exploring their implications in cosmological contexts. Particular attention has been devoted to non-linear extensions formulated through arbitrary functions of the torsion scalar $f({\mathbb T})$ or the non-metricity  scalar $f({\mathbb Q})$ (see \cite{Heisenberg:2023lru,Cai:2015emx} for reviews). Another prominent line of research involves constructing all possible independent quadratic polynomials of the torsion and non-metricity tensors, giving rise to the so-called New and Newer General Relativity \cite{PhysRevD.19.3524,BeltranJimenez:2017tkd}. Upon relaxing the assumption of parity conservation, there is one parity-odd quadratic combination of the non-metricity tensor \cite{Iosifidis:2018zwo} which has been  studied in details in the context of gravitational wave phenomenology  \cite{Conroy:2019ibo,Chen:2022wtz}. Linear perturbation analyses around Minkowski spacetime have shown that Newer General Relativity is generally prone to ghost instabilities \cite{Heisenberg:2023lru,BeltranJimenez:2018vdo}, thereby  requiring degeneracy conditions on the free parameters to avoid them. These conditions correspond to enforcing linear transverse diffeomorphism (TDiff) invariance. Remarkably, as shown in \cite{Bello-Morales:2024vqk}, this linear TDiff symmetry can be promoted to the non-linear level, giving rise to a stable subclass of theories extending beyond STEGR. Moreover, this class exhibits close connections with scalar-tensor theories and with the family of transverse gravity models \cite{Alvarez:2006uu,Alvarez:2009ga,Blas:2011ac,Bello-Morales:2023btf}. 

However, linear perturbations around highly symmetric backgrounds may misrepresent the true number of physical degrees of freedom (DOFs), either underestimating or, in certain cases, overestimating them \cite{Ganz:2022iiv}.  Therefore, a non-linear analysis is crucial to explore these models. 
The most widely used method for determining the number of DOFs is the Dirac-Bergmann algorithm \cite{Dirac:1958sq,dirac2001lectures,anderson1951constraints}, formulated within the Hamiltonian framework. 
Nonetheless, the application of this algorithm has led to conflicting results in the literature, both for $f({\mathbb T})$ \cite{Li:2011rn,Ferraro:2018tpu,Blagojevic:2020dyq,Blixt:2020ekl} and $f({\mathbb Q})$ \cite{DAmbrosio:2023asf,Hu:2022anq,Tomonari:2023wcs} theories. 

In \cite{DAmbrosio:2023asf} the authors linked the problem  to the emergence of partial differential equations  (PDEs) for  the Lagrange multipliers - rather than linear algebraic equations - which causes a breakdown of the Dirac-Bergmann algorithm. On the other hand, in \cite{Heisenberg:2025fxc} the author proposed a novel approach to count the number of DOFs in these models,  based on a work of Einstein \cite{Einstein1955}, the Cartan-Kuranishi algorithm \cite{Cartan1930,Cartan1945,Kuranishi1957} and extending the earlier works by Seiler \cite{Seiler:1995ne,Seiler:1995Fa,Seiler2000,Seiler2010}. In this framework, instead of relying on the Hamiltonian formalism, the analysis is carried out directly at the level of the field equations, which are rendered involutive by systematically identifying hidden integrability conditions via the Cartan–Kuranishi procedure.  Interestingly, in \cite{Seiler:1995ne} it has been discussed that, for finite-dimensional systems, the Dirac-Bergmann algorithm operates on a conceptually similar basis, by turning  the Hamiltonian equations into an involutive system. 

In this work, we establish a comprehensive analysis of Newer General Relativity extended by the distinctive quadratic parity-violating operator. We begin by revisiting the structure of the theory within this enlarged framework, emphasizing the novel features introduced by the parity-violating term. The paper is organized as follows. In Section \ref{sec:Preliminaries}, we review the essential elements of non-metricity theories. In Section \ref{sec:Perturbation} we carry out a detailed analysis of linear perturbations on a cosmological (FLRW) background and show that the class of theories under consideration generically exhibits ghost instabilities. There are, however, two notable exceptions. The first is STEGR augmented by a parity-violating operator. The second is the aforementioned TDiff-invariant gravity subclass, also involving the parity-violating term, that remains ghost-free at the linear level but, in contrast to STEGR, propagates an additional scalar DOF in the gravitational sector alongside the two usual tensor modes. Crucially, the inclusion of the parity-violating operator breaks the non-linear TDiff symmetry, thereby driving the system into a strong-coupling regime around FLRW. As a result, additional DOFs are expected to become dynamical beyond linear order. Thus, although these subclasses avoid linear ghosts, the symmetry breaking induced by the parity-violating term signals a breakdown of the perturbative description and the emergence of further non-linear dynamical modes. Motivated by these findings, in Section \ref{sec:Non_linear_DOF} we undertake, for the first time, a systematic non-linear determination of the total number of propagating DOFs. To isolate the effect of parity violation on the dynamical content of the theory, we focus on the case of STEGR supplemented by the parity-violating operator. This choice aligns with the standard setup employed in parity-violating gravity models; moreover, the analysis straightforwardly extends to the full TDiff-invariant class, and such generalizations do not alter the final conclusions regarding the number of propagating DOFs. We then compare the Hamiltonian analysis via the Dirac-Bergmann algorithm with the newly proposed method \cite{Heisenberg:2025fxc} based on the Cartan-Kuranishi algorithm. Through this joint approach, we definitively show that the theory generally propagates eight DOFs. 

We use the mostly positive signature $(-,+,+,+)$  and work in units where $M_{\rm pl}^2=c^4/(8\pi G)=1$ and $c=\hbar =1$.

\section{Review of Non-Metricity Theories}
\label{sec:Preliminaries}

Let us briefly review the essential elements of metric–affine geometry relevant for the discussion of quadratic non-metricity theories (see \cite{Heisenberg:2023lru} for a comprehensive review). In this general formulation, the metric $g_{\mu\nu}$ and the affine connection $\tensor{\Gamma}{^\lambda_{\mu\nu}}$ are treated as independent dynamical fields, without assuming any a priori relation between them. The connection is further restricted to be flat and torsion-free, so that curvature and torsion vanish identically, \textit{i.e.}
\begin{align}
    \tensor{R}{^\rho_\sigma_\mu_\nu} =& \,\,\partial_\mu \tensor{\Gamma}{^\rho_\sigma_\nu}- \partial_\nu \tensor{\Gamma}{^\rho_\sigma_\mu} + \tensor{\Gamma}{^\rho_\lambda_\mu} \tensor{\Gamma}{^\lambda_\sigma_\nu} - \tensor{\Gamma}{^\rho_\lambda_\nu} \tensor{\Gamma}{^\lambda_\sigma_\mu} \overset{!}{=}  0~, \\
    \tensor{T}{^\rho_\mu_\nu} =& \,\,\tensor{\Gamma}{^\rho_\mu_\nu} - \tensor{\Gamma}{^\rho_\nu_\mu}\overset{!}{=}0~.
\end{align}
The two conditions can  be solved for the affine connection leading to \cite{Adak:2008gd,BeltranJimenez:2017tkd}
\begin{align}
    \tensor{\Gamma}{^\rho_\mu_\nu} = \frac{\partial x^\rho}{\partial \xi^\lambda } \partial_\mu \partial_\nu \xi^\lambda~,
\end{align}
where $\xi^\lambda$ are arbitrary functions of spacetime coordinates $x^\alpha$ with the condition that $\partial  x^\alpha/\partial \xi^\lambda$ is non-degenerate. 
By gauge fixing $\xi^\lambda=x^\lambda$, the affine connection trivializes $\tensor{\Gamma}{^\lambda_\mu_\nu} =0$. This is known as the coincident gauge \cite{Adak:2008gd,BeltranJimenez:2017tkd}. All the dynamical DOFs are shifted into the metric, which leads to a trivial upper bound of 10 DOFs for any non-metricity theory. Note that, more generally, the coincident gauge is not spoiled by any linear coordinate transformation $x^\alpha \rightarrow \tensor{M}{^\alpha_\beta} x^\beta + c^\alpha$ with $\tensor{M}{^\alpha_\beta}$ and $c^\alpha$ being constant \cite{BeltranJimenez:2018vdo,BeltranJimenez:2022azb,Heisenberg:2025fxc}.

In metric-affine gravity, the dynamics of the metric are described by the non-metricity tensor
\begin{align}
\label{eq:Non-metrcity-tensor}
    Q_{\alpha\mu\nu} = \nabla_\alpha g_{\mu\nu} = \partial_\alpha g_{\mu\nu} -2 \tensor{\Gamma}{^\lambda_\alpha_(_\mu} g_{\nu)\lambda}~.
\end{align}
Up to quadratic order in the non-metricity tensor there are six distinct scalars from which one breaks parity. Thus, the most general action up to second order is given by \cite{Iosifidis:2018zwo}
\begin{align}
\label{action}
    {\mathcal S} = \int  \md^4x\, \sqrt{-g} \Big[ c_1Q_{\alpha\mu\nu}Q^{\alpha\mu\nu}+c_2Q_{\mu\alpha\nu}Q^{\alpha\mu\nu}+c_3Q_{\mu}Q^{\mu}+c_4\bar{Q}_{\mu}\bar{Q}^{\mu}+c_5Q_{\mu}\bar{Q}^{\mu} + c_{\rm  PV}  \epsilon^{\mu\nu\rho\sigma}Q_{\mu\nu\lambda}Q_{\rho\sigma}^{\,\,\,\,\,\,\lambda} \Big]~,
\end{align}
where
\begin{align}
    Q_\mu = \tensor{Q}{_\mu_\nu^\nu}, \qquad \bar Q_\mu = \tensor{Q}{^\alpha_\alpha_\mu}~,
\end{align}
and $c_{\rm  PV}$, $c_i$ for $i=\{1,2,3,4,5\}$ are constant.

There is one particular  relevant combination of the quadratic tensors, which is known as the non-metricity scalar ${\mathbb Q}$
\begin{align}
\label{Q}
    {\mathbb Q}  =& -  \frac{1}{4} Q_{\alpha\mu\nu}Q^{\alpha\mu\nu}+ \frac{1}{2} Q_{\mu\alpha\nu}Q^{\alpha\mu\nu}+\frac{1}{4} Q_{\mu}Q^{\mu}- \frac{1}{2} \bar{Q}_{\mu}\bar{Q}^{\mu} \nonumber \\
    =& {\cal R} + {\hat \nabla}_\mu \left( Q^\mu - \bar Q^\mu \right)~,
\end{align}
where ${\cal R}$ is the Ricci scalar with respect to the Levi-Civita connection and $\hat \nabla$  the Levi-Civita covariant derivative. Since this particular combination is equivalent to the Ricci scalar up to boundary terms, the corresponding theory reduces to GR \cite{BeltranJimenez:2017tkd}.

\section{Cosmological Perturbation Theory}
\label{sec:Perturbation}

In this section, we discuss the linear perturbations for the parity-violating quadratic non-metricity gravity theory Eq. \eqref{action} 
on the flat Friedmann-Lemaître-Robertson-Walker (FLRW) background.

\subsection{Background equation}

For simplicity we consider the flat FLRW metric
\begin{align}
    \md s^2 = - N^2 \md t^2 + a^2 \md x^i\md x^j \delta_{ij}~,
\end{align}
where $a(t)$ is the scale factor and $N(t)$ is the lapse  function. 

The most general connection compatible with the cosmological  principle has been derived  in  \cite{Hohmann:2021ast,DAmbrosio:2021pnd}, where it has been shown that it splits into  three different classes for a spatially flat FLRW background. Note that the coincident gauge is consistent with the diagonal ansatz of the metric corresponding to one of the three classes. Moreover, by performing coordinate transformation, the other two classes can also be brought into the coincident gauge. However, in that case the metric is no longer diagonal \cite{Hohmann:2021ast,Jensko:2024bee}. For simplicity, we only consider the diagonal metric ansatz and the coincident gauge $\tensor{\Gamma}{^\alpha_\mu_\nu}=0$, limiting  our discussion to one of the three classes.

Using the diagonal metric ansatz and the coincident gauge, the background Lagrangian, following from Eqs. \eqref{action}, simplifies to
\begin{equation}
\label{bgLagrangian}
    \mathcal L_{\rm BG}=-4 a^3\left[A\frac{\dot N^2}{N^3}+3B\frac{H\dot N}{N^2}+3C\frac{H^2}{N}\right]~,
\end{equation}
 where \(H=H(t)\equiv \dot a/a\) denotes the Hubble parameter and for notational convenience, we have defined the constants
\begin{equation}
A \equiv c_1+c_2+c_3+c_4+c_5~\,\\,\,\,\,\,\,\,\,\,\,\,\,
B \equiv 2c_3+c_5~\,\\,\,\,\,\,\,\,\,\,\,\,\,
C \equiv c_1+3c_3~.
\end{equation}
Note that for $A\neq0$ and $B\neq 0$ in general  there are two dynamical DOFs at the background: $N$ and $a$. In particular, we cannot fix $N=1$ as the  gauge freedom has been already fixed by choosing the coincident gauge. It is noteworthy that, at the mini-superspace level, the theory still propagates two DOFs even after imposing the non-linear TDiff conditions $c_5=2 c_1$, $c_4=0$ and $c_2 = - 2 c_1$ (see \cite{Bello-Morales:2023btf} for a detailed discussion of background solutions within this class of models). The only exception is the STEGR limit $c_3 =-c_1$, in which case no dynamical DOF remains in mini-superspace.

Varying the Lagrangian \eqref{bgLagrangian} with respect to the lapse function and the scale factor yields the following equations of motion
\begin{align}
    &A\left(3\frac{\dot N^2}{N^4}-2\frac{\ddot N}{N^3}-6\frac{H\dot N}{N^3}\right)-3B\left(\frac{\dot H}{N^2}+3\frac{H^2}{N^2}\right)-3C\frac{H^2}{N^2}=0~, \\&
    A\frac{\dot N^2}{N^3}+B\left(2\frac{\dot N^2}{N^3}-\frac{\ddot N}{N^2}\right)+C\left(2\frac{H\dot N}{N^2}-2\frac{\dot H}{N}-3\frac{H^2}{N}\right)=0~.
\end{align}
For a detailed discussion of the background equation in Newer GR in the other classes see \cite{Hohmann:2021ast} as the parity-violating operator does not  impact the background equations.

\subsection{Linear perturbations}
At linear order, the metric can be decomposed as \(g_{\mu\nu} = \bar{g}_{\mu\nu} + \delta g_{\mu\nu}\), where \(\bar{g}_{\mu\nu}\) is the unperturbed background metric and \(\delta g_{\mu\nu}\) denotes the metric perturbations. According to the standard SVT decomposition, these perturbations can be systematically written as
\begin{align}
\label{eq12}
    &\delta g_{00}=-2N^2\alpha~,\\&
    \label{eq13}
    \delta g_{0i}=a^2\mathcal B_i \equiv a^2\left(\partial_i \beta + B_i \right)~,\\&
    \delta g_{ij}=2a^2 \mathcal C_{ij} \equiv 2a^2\left(-\delta_{ij}\psi+ \frac{1}{2}\partial_i\partial_j E+\partial_{(i}C_{j)}+\frac{1}{2}h_{ij}\right)~,
\label{eq14}
\end{align}
 where \(\alpha\), \(\beta\), \(\psi\) and \(E\) are scalar perturbations, \(B_i\) and \(C_i\) are vector perturbations, and \(h_{ij}\) is a tensor perturbation. In particular, vector perturbations are divergence-free (\(\partial^iB_i=\partial^iC_i=0\)), while the tensor perturbation is both transverse (\(\partial^ih_{ij}=0\)) and traceless (\(h^i_{\,\,i}=0\)). 

 On the other hand, the perturbations of the connection are given by
 \begin{align}
     \delta \tensor{\Gamma}{^\lambda_\mu_\nu} = \partial_\nu \partial_\mu \zeta^\lambda ~,
 \end{align}
 where $\zeta^\lambda$ can be decomposed into two scalar and vector modes as usual $\zeta_0 = a F$, $\zeta_k = a( \partial_k J + \hat J_k )$  \cite{Heisenberg:2023tho,Heisenberg:2023wgk}. 
In the following, we will use the coincident gauge ($F=0$, $J=0$ and $\hat J_k=0$) so that all the dynamical fields are encoded by the metric perturbations. 
Finally, prior to proceeding with the perturbative analysis, it is convenient to re-express the parity-violating contribution in terms of the Levi-Civita symbol \(\tilde\epsilon^{\mu\nu\rho\sigma}\) which is related to the Levi-Civita tensor (a tensor density of weight \(-1\)) by \(\epsilon^{\mu\nu\rho\sigma}={\rm sgn}(g)\tilde\epsilon^{\mu\nu\rho\sigma}/\sqrt{-g}\). 

In the next subsections, we focus on the scalar, vector and tensor sector separately.

\subsubsection{Tensor sector}
The tensor Lagrangian, which follows from the decomposition of Eq. \eqref{action}, is given by
\begin{equation}
    \mathcal L^{T}=c_1\left(Na\partial_kh_{ij}\partial^kh^{ij}-\frac{a^3}{N}\dot h_{ij} \dot h^{ij}\right)-2c_{PV}a^2\tilde\epsilon^{ijk}\dot h_{il}\partial_jh_k^{\,\,\,l}~.
\end{equation}
It follows immediately that, in order to avoid ghost modes (assuming \(N>0\) and \(a>0\)), the condition
\begin{equation}
\label{Tcond}
    c_1<0
\end{equation}
must be imposed, recovering two tensor DOFs. The tensor equations of motions read
\begin{equation}
\label{Teom}
    \ddot h_{ij}+\left(3H-\frac{\dot N}{N}\right)\dot h_{ij}-\frac{N^2}{a^2}\partial_k\partial^k h_{ij}+\frac{2c_{PV}HN}{c_1a}\tilde\epsilon_{ilk}\partial^l h^k_{\,\,\,j}=0~.
\end{equation}
Note that in comparison to the Chern-Simons  coupling \cite{Lue:1998mq,PhysRevD.68.104012,Alexander:2009tp}, the tensor sector is stable in the UV-limit $k \rightarrow \infty$ and one does not need to impose a Cut-Off scale. Instead, the parity violation leads to a tachyonic-like instability for scales
\begin{align}
    k < \frac{2c_{PV}aH}{c_1N}~.
\end{align}
For a more detailed analysis we refer to \cite{Conroy:2019ibo,Chen:2022wtz,Jenks:2023pmk}. 

\subsubsection{Vector  sector}

We now turn to the vector sector, associated with the divergence-free vectors \(B_i\) and \(C_i\). The vector Lagrangian is given by
\begin{align}
    \mathcal L^V&=(2c_1+c_2+c_4)\frac{a^3}{N}\left[\dot B_i\dot B^i+2\left(\frac{\dot N}{N}-H\right)B_i\dot B^i+\left(\frac{\dot N}{N}-H\right)^2B_iB^i\right] \nonumber \\ &-2c_1Na\partial_jB_i\partial^jB^i-2a^2(c_2+c_4)\dot B_i\partial_j\partial^jC^i+2c_4a^2\left(H-\frac{\dot N}{N}\right)B_i\partial_j\partial^jC^i\notag\\&+(2c_1+c_2+c_4)Na\partial_j\partial^jC_i\partial_k\partial^kC^i-2c_1\frac{a^3}{N}\partial_j\dot C_i\partial^j\dot C^i+2c_{PV}a^2\epsilon_{ikl}\partial_j\dot C^l\partial^k\partial^jC^i~.
\end{align}
 To evaluate the kinetic matrix \(K^V_{ij}\) we move to the Fourier space, obtaining
\begin{equation}
    K_{ij}^V=\frac{a^3}{N}
    \begin{pmatrix}
2c_1+c_2+c_4 & 0 \\
0 & 2c_1k^2
\end{pmatrix}~.
\end{equation}
Taking into account the condition \eqref{Tcond} that arises from the tensor sector, the issue is manifest: the eigenvalue associated with \(C_i\) is found to be negative, signaling the presence of a ghost. Hence, one must impose that the determinant of the kinetic matrix vanishes, thus obtaining the condition
\begin{align}
\label{Vcond}
    2c_1+c_2+c_4=0~,
\end{align}
which corresponds to the linear TDiff-invariance \cite{ALVAREZ2006148}.
This constraint makes \(B_i\) manifestly non-dynamical and simultaneously cures the pathology associated with \(C_i\). Indeed, solving the equations of motion for \(B_i\) gives
\begin{equation}
    B_i=\frac{a}{N}\left(2HC_i+\dot C_i\right)-\frac{c_4a}{2c_1N}\left(H-\frac{\dot N}{N}\right)C_i~.
\end{equation}
Substituting this solution back into \(\mathcal L^V\) renders \(C_i\) non-dynamical as well. Consequently, the theory possesses no healthy propagating vector DOFs.

\subsubsection{Scalar sector}
 Finally, we turn to the dynamical contributions arising from the scalar sector, associated with the scalar perturbations \(\alpha\), \(\beta\), \(E\) and \(\psi\). Because the full scalar Lagrangian (see Appendix \ref{scalarLagrangian}) is quite cumbersome, we display only the kinetic contributions, which are the only terms relevant for the present stability analysis.
\begin{align}
\label{SLagrangian}
\mathcal L^S_{kin}&=\frac{a^3}{N}\bigg[4(c_{1}-c_{3}-c_{5})\,\dot{\alpha}^{2}-(c_{1}+c_{3})\left(\partial_i \partial^i\dot{E}\right)^{2} -12(c_{1}+3 c_{3})\,\dot{\psi}^{2}\notag\\ &-2(2 c_{3}+c_{5})\dot{\alpha}\,\partial_i \partial^i\dot{E}+12(2 c_{3}+c_{5})\,\dot{\alpha}\,\dot{\psi}  +4(c_{1}+3 c_{3})\partial_i \partial^i\dot{E}\,\dot{\psi}  \bigg]~.
\end{align}
 Note that the constraint \eqref{Vcond} coming from the vector sector has already been imposed and that the background field equations were used to simplify terms such as \(\dot H\). 

Rather than working directly with $E$, it is convenient to introduce a new variable $V$, defined as
\begin{align}
    V \equiv \delta g = 2 \alpha + \partial^2 E - 6 \psi~.
\end{align} 
As is apparent from Eq. \eqref{SLagrangian}, \(\beta\) is manifestly non-dynamical. The equation of motion for \(\beta\) then takes the form
\begin{align}
    &(2c_1-c_5) \dot V - 8 c_1 \dot \psi - 8 (c_1 + c_4) H \alpha + 4 \left(3  H - \frac{\dot N}{N} \right)c_4 \psi \nonumber \\
    & - \left( (6 c_3 + 2 c_4 + 3 c_5) H + (2 c_3 + c_5) \frac{\dot N}{N} \right) V=0~.
    \label{eq29}
\end{align}
Solving the expression for $\alpha$ and substituting the result back into the Lagrangian Eq. \eqref{SLagrangian} we obtain, after performing several integrations by parts, 
\begin{align}
\label{eq:Lagrangian_Two_scalar}
    \mathcal L^S = a^3 N \Big[ \frac{K_{11}}{2 N^2} \dot V^2 + \frac{K_{12}}{N^2} \dot V \dot \psi + \frac{K_{22}}{2 N^2} \dot \psi^2 + F (\dot \psi V - \dot V \psi ) + \frac{1}{2} G_{11} V^2 + G_{12} V \psi + \frac{1}{2} G_{22} \psi^2   \Big]~.
\end{align}
The explicit forms of $K_{ij}$, $F$ and $G_{ij}$ are rather cumbersome and offer limited conceptual insight. For the purposes of the subsequent discussion, we relegate the expressions for the kinetic terms to the Appendix \ref{subsec:Kinetic_terms}.  
We observe that for generic values of $c_3$, $c_4$ and $c_5$, the theory propagates two dynamical scalar DOFs. This stands in contrast to the analysis of linear perturbations around Minkowski spacetime, where only one scalar mode remains dynamical once the linear TDiff condition \eqref{Vcond} is imposed. Consequently, these models are generically strongly coupled around Minkowski, which is a well-known issue in non-metricity theories \cite{Gomes:2023tur}.

Imposing condition \eqref{Tcond} to ensure the stability of tensor perturbations, the kinetic matrix becomes degenerate for a general background evolution if and only if
\begin{align}
\label{eq:Condition_non_linear_TDIFF}
    c_5 = 2 c_1~, \qquad c_4 =0.
\end{align}
Interestingly, these two conditions guarantee that, aside from the parity-violating term, which does not affect the scalar perturbations, the non-linear TDiff symmetry is restored \cite{Bello-Morales:2024vqk}. 

Upon imposing the conditions \eqref{eq:Condition_non_linear_TDIFF}, the kinetic matrix becomes degenerate, allowing us to define a new variable
\begin{align}
    Y \equiv V - 2 \left( 3 + \frac{\dot N }{H N} \right) \psi~,
\end{align}
such that $\psi$ drops out of the Lagrangian, which then takes the final form 
\begin{align}
    {\cal L}^S = a^3 N \Big[ - \frac{ c_1 + c_3 }{N^2} \dot Y^2 - \left( - (c_1 + c_3) \frac{k^2}{a^2} + M^2 \right) Y^2 \Big]~,
\end{align}
where 
\begin{align}
    M^2 = \frac{( (2 c_1 + 3 c_3) H N + (c_1 + c_3) \dot N)^2 }{2 c_1^2 H^2 N^6 } \left( 3 (c_1 + 3 c_3) H^2 N^2  + 6 ( c_1 + c_3) H N \dot N + (c_1+ c_3) \dot N^2  \right) ~.
\end{align}
To ensure the stability of the perturbations, we require
\begin{align}
    c_3 < - c_1 ~,
\end{align}
which is consistent with \cite{Bello-Morales:2024vqk}, where this bound can be directly inferred from the scalar-tensor formulation. 
As expected, for $c_3=-c_1$ the scalar DOF vanishes, recovering STEGR. It is worth emphasizing that, as noted above, the parity-violating operator does not influence the scalar sector; however, it explicitly breaks the non-linear TDiff symmetry. Consequently, at higher orders or around less symmetric backgrounds, additional DOFs are expected to become dynamical, indicating strong coupling around highly symmetric configurations such as Minkowski or FLRW spacetimes.

\section{Non-linear degrees of freedom}
\label{sec:Non_linear_DOF}

To investigate the possible emergence of additional DOFs in the presence of the parity-violating operator, we focus on STEGR, corresponding to $2c_1=- c_2 = -2c_3= c_5 =-1/4$ and $c_4=0$ supplemented by the parity-violation term. The action then simplifies to
\begin{align}
\label{eq:GR_Parity_violation}
    {\mathcal S} = \int \md^4x\, \sqrt{-g} \Big[ \frac{1}{2} {\mathbb Q} + c_{\rm PV} \epsilon^{\mu\nu\rho\sigma} Q_{\mu\nu\alpha} \tensor{Q}{_\rho_\sigma^\alpha} \Big]~.
\end{align}
Nevertheless, the general conclusions remain valid upon extending the ansatz to include an arbitrary $c_3$. 

As in the previous section, we will employ the coincident gauge to shift all the DOFs from the connection to the metric. Note that for $c_{\rm PV}=0$ the gauge freedom is restored due to the specific form of ${\mathbb Q}$ Eq. \eqref{Q} and one recovers GR as discussed in details in \cite{Nester:1998mp,BeltranJimenez:2017tkd,DAmbrosio:2020nqu,BeltranJimenez:2019esp}.

In this section,  we determine the number of propagating DOFs for the parity violation extension of STEGR Eq. \eqref{eq:GR_Parity_violation} by performing a Hamiltonian analysis using the Dirac-Bergmann algorithm, and by applying an alternative approach \cite{Seiler:1995ne, Heisenberg:2025fxc} based on the Cartan-Kuranishi algorithm. 

\subsection{ADM decomposition}

In order to carry out the Hamiltonian analysis, we first perform the ADM decomposition, in which the metric is given by
\begin{align}
    \md s^2 = - N^2 \md t^2 + \gamma_{ij} \left( N^i \md t + \md x^i \right) \left( N^j \md t+ \md x^j \right)~,
\end{align}
where $N$, $N^i$ and $\gamma_{ij}$ are the lapse function, the shift vector and the 3-dimensional induced metric on the hypersurface of constant time, respectively. 

Following \cite{DAmbrosio:2020nqu}, the non-metricity tensor Eq. \eqref{eq:Non-metrcity-tensor} in the coincident gauge can be decomposed as
\begin{align}
    n^\alpha Q_{\alpha j k} = & \,\,\frac{\gamma_{ik} \partial_j N^k + \gamma_{ij} \partial_k N^i }{N} + 2 K_{jk}, \\
    n^\alpha Q_{i\alpha k} =& \,\,\frac{1}{N} \gamma_{jk} \partial_i N^j, \\
    n^\alpha n^\beta Q_{\alpha\beta k} = & \,\,\frac{\gamma_{jk}}{N^2} \left( \dot N^j - N^i \partial_i N^j\right) \\
     n^\alpha n^\beta Q_{i \alpha\beta} =& - \frac{2 \partial_i N}{N}, \\
     n^\alpha n^\beta n^\gamma Q_{\alpha\beta\gamma} =& \,\,\frac{2}{N^2} \left( N^k \partial_k N - \dot N \right)~,
\end{align}
where $n^\mu=(1/N,-N^i/N)$ is the normal vector to the hypersurface of constant time and $K_{ij}$ is the extrinsic curvature given by
\begin{align}
    K_{ij} = \frac{1}{2N} \left( \dot \gamma_{ij} + {\cal D}_i N_j + {\cal D}_j N_i \right)~,
\end{align}
with ${\cal D}_i$ being the Levi-Civita covariant derivative with respect to the induced metric. Note that the presence of the covariant derivative should not be confused with original non-metric-compatible connection which has been trivialized by choosing the coincident gauge. 

In \cite{DAmbrosio:2020nqu} it has been discussed that up to boundary terms the STEGR can be brought into the form
\begin{align}
    {\mathcal S}_{\rm GR} = \frac{1}{2} \int \md^4x\, \sqrt{\gamma} N \left( ^{(3)}{\cal R} + K_{ij} K^{ij} - K^2 \right)~,
\end{align}
where $^{(3)}{\cal R}$ is the 3-dim intrinsic curvature with respect to the Levi-Civita connection. On the other hand, the parity-violating contribution, corresponding to the last term in Eq. \eqref{eq:GR_Parity_violation}, yields
\begin{align}
    {\mathcal S}_{\rm PV} = 2c_{\rm PV} \int \md^4x\, \sqrt{\gamma} N \epsilon^{0ijk} & \Big[ \gamma^{mn} \partial_j \gamma_{kn} \left( \frac{2}{N} \gamma_{lm} \partial_i N^l + \frac{1}{N} \gamma_{li} \partial_m N^l + 2 K_{im}   \right) \nonumber \\
    & - \frac{1}{N^2} \left( \frac{1}{N} \gamma_{li} \left( \dot N^l - N^r \partial_r N^l \right) - 2 \partial_i N  \right) \gamma_{mk} \partial_j N^m  \Big]~.
    \label{eq:ADM_Parity_Violation}
\end{align}
In the following, it is useful to introduce the 3-dim Levi-Civita tensor $N \epsilon^{0ijk} =  \epsilon^{ijk} $. It is straightforward to verify that the parity-violating term explicitly breaks the diffeomorphism invariance of GR. As a preliminary estimate, one might therefore expect at least six propagating DOFs at the fully non-linear level, corresponding to the loss of gauge symmetries.

On the other hand, the kinetic matrix remains unaffected, since the parity-violating contributions are only linear in the time derivatives. Therefore, there will be four primary constraints in the Hamiltonian analysis leading to an upper bound of eight DOFs. As the lapse is non-dynamical we  can derive the constraint equation to the lapse function  which leads to an algebraic fourth order polynomial of the lapse function 
\begin{align}
    N^4 \frac{\delta \mathcal L}{\delta N} = f(N, \gamma_{ij}, N^l, \dot \gamma_{ij}) + 6\sqrt{\gamma} c_{\rm PV} \tensor{\epsilon}{^i^j^k}\gamma_{li}\dot N^l \gamma_{mk} \partial_j N^m  =0~,
\end{align}
where $\mathcal L$ is the Lagrangian related to the action Eq. \eqref{eq:GR_Parity_violation}.
Solving the constraint equation for the lapse function and substituting it back into the action introduces non-linear kinetic terms for the shift vector. As a result, some or all components of the shift may become dynamical. One could try to rederive the updated kinetic matrix to get a better understanding of the DOFs. However, solving the constraint equation for the lapse function is highly  non-trivial and instead we will  employ the Dirac-Bergmann algorithm. 

\subsection{Dirac-Bergmann algorithm}

The canonical conjugate momentum of the metric is given by
\begin{align}
    \pi^{ij} =& \,\,\frac{1}{2} \sqrt{\gamma} \left(  -K \gamma^{ij} + K^{ij} \right) + 2c_{\rm PV} \sqrt{\gamma} \epsilon^{0rlk} \gamma^{mn} \partial_l \gamma_{kn}  \tensor{\delta}{_r^{(i}} \tensor{\delta}{_m^{j)}} \nonumber \\
     =& \,\,\frac{1}{2} \sqrt{\gamma} \left(  -K \gamma^{ij} + K^{ij} \right) + 2c_{\rm PV} \sqrt{\gamma} \frac{1}{N} \tensor{\epsilon}{^l^k^(^i} \gamma^{j)n}\partial_l \gamma_{kn}~. 
\end{align}
Note that the trace of the second term vanishes due the symmetries of the Levi-Civita tensor and, therefore, we can invert the expression as
\begin{align}
    K^{ij} = \frac{2}{\sqrt{\gamma}} \left( \pi^{ij} - \frac{1}{2} \tensor{\pi}{^k_k} \gamma^{ij} - 2 c_{\rm PV} \sqrt{\gamma} \frac{1}{N} \epsilon^{lk(i} \gamma^{j)n} \partial_l \gamma_{kn}   \right)~.
\end{align}
On the other hand, the momenta for the shift and the lapse are constraints
\begin{align}
    \pi_N  \approx & \,\,0,\\
    C_r \equiv& \,\,\pi_r +  2c_{\rm PV} \sqrt{\gamma} \epsilon^{ijk} \frac{1}{N^3} \gamma_{ri} \gamma_{mk} \partial_j N^m \approx 0~,
\end{align}
where $\approx $ denotes weak equalities on the constrained hypersurface. 

Before going into details of the calculations, we can already note that the two constraints do not commute with each other, instead
\begin{align}
    \{ \pi_N, C_r\} = 6 c_{\rm PV} \sqrt{\gamma} \tensor{\epsilon}{_r^j_m} \frac{1}{N^4} \partial_j N^m \approx - \frac{3}{N} \pi_r~,
\end{align}
where,  in the second step, we used the constraint $C_r$ to simplify the expression. Moreover, $C_r$ will in general also not commute with itself as it depends on derivatives of the shift vector. For the computation of the commutator $C_r$ with itself it is convenient to note that any terms algebraic in  the shift vector and its canonical momentum drop out and, therefore, one can replace  the partial derivative with the Levi-Civita covariant derivative, thus leading to
\begin{align}
    \big \{ C_r ,  \int \md^3x u^l C_l  \big \}=  &  \,\,\sqrt{\gamma}  {\cal D}_j \left( u^l 2c_{\rm PV} \tensor{\epsilon}{_l^j_r} \frac{1}{N^3}  \right)  - \sqrt{\gamma} {\cal D}_j \left(   2c_{\rm PV} \tensor{\epsilon}{_r^j_l} u^l \frac{1}{N^3} \right) \nonumber \\
  = & \,\,4 c_{\rm PV} {\cal D}_j \left(  \sqrt{\gamma} u^l \tensor{\epsilon}{_l^j_r} \frac{1}{N^3}  \right)~,
\end{align}
where we have smeared over one of the constraints to avoid Dirac-Delta functions. 

The total Hamiltonian is given by
\begin{align}
    H_T =\int \md^3x\,\Big[ {\cal H}(N,N^k,\gamma_{ij},\pi^{ij}) + u_N \pi_N + u^r C_r \Big]~,
\end{align}
where 
\begin{align}
    {\cal H} = & N \left( \frac{2}{\sqrt{\gamma}} \left( \pi^{ij} \pi_{ij} - \frac{1}{2} \pi^2 \right) - \frac{1}{2} \sqrt{\gamma}\, ^{(3)}{\cal R}  \right) - 2 \pi^{ij} {\cal D}_i N_j - 2 \sqrt{\gamma} \epsilon^{ijk} c_{\rm PV} \Big[ \gamma^{mn} \partial_j \gamma_{kn}  \Big( \frac{2}{N} \gamma_{lm} \partial_i N^l \nonumber \\
    &   + \frac{1}{N} \gamma_{li} \partial_m N^l - \frac{8 c_{\rm PV}}{N} \tensor{\epsilon}{^l^r_(_i} \tensor{\gamma}{_{m)}^s} \partial_l \gamma_{rs}      \Big)+ \frac{1}{N^2} \left(  \frac{1}{N} \gamma_{li}   N^r \partial_r N^l  + 2 \partial_i N  \right) \gamma_{mk} \partial_j N^m  \Big]~.
\end{align}
The conservation of the constraints leads to four conditions on the Lagrange multipliers  $u_N$ and $u^k$, \textit{i.e.}
\begin{align}
    &\{\pi_N , {\cal H} \} - \frac{3}{N}  u^r \pi_r \approx 0,\label{eq:Lagrange_Multiplier_1} \\
    &  \{C_r , {\cal H} \} + \frac{3}{N} u_N \pi_r + 4 c_{\rm PV} {\cal D}_j \left(  \sqrt{\gamma} u^l \tensor{\epsilon}{_l^j_r} \frac{1}{N^3}  \right)\approx 0~.  \label{eq:Lagrange_Multiplier_2}
\end{align}
We can note that, while the first equation is an algebraic constraint, the other three equations are  PDEs. This is in agreement with the claim that, in general,  non-metricity theories will have non-algebraic equations for the Lagrange  multipliers leading to a breakdown of the conventional Dirac-Bergmann algorithm, as  it is based on algebraic conditions \cite{DAmbrosio:2023asf}. 

In Appendix \ref{app:Lagrange_mulitplier}, we analyze in detail the PDEs governing the Lagrange multipliers and demonstrate that, under appropriate spatial boundary conditions, they admit a unique solution. This behavior is conceptually analogous to what occurs in $f({\mathbb T})$ gravity \cite{Li:2011rn,Blagojevic:2020dyq,Tomonari:2023wcs}, where one of the Lagrange multipliers is likewise uniquely determined - only once suitable boundary conditions are imposed (see Appendix \ref{app:Lagrange_mulitplier} for further discussion). In contrast, the situation differs markedly from the results of \cite{DAmbrosio:2023asf} for $f({\mathbb Q})$ gravity, where the corresponding PDEs never yield unique solutions for the Lagrange multipliers. In that case, the loss of uniqueness indicates a breakdown of the Dirac-Bergmann procedure and the appearance of additional constraints. A related perspective is provided in \cite{Seiler:1995ne}, where it is shown that, in field theories, treating the constraints purely algebraically and performing a  prolongation only in the time direction, the Dirac-Bergmann algorithm can overlook integrability conditions, thereby leading to non-uniquely determined Lagrange multipliers.

While the Dirac-Bergmann algorithm is a priori not well defined for PDEs, the unique solution for proper boundary conditions may be a proof that all the Lagrange multipliers are properly fixed and, therefore, there are only 4 second-class constraints. The number of DOFs is given by
\begin{align}
    \# \textrm{DOFs} = \frac{ N_C - 2 N_F - N_S  }{2}  = \frac{20-0-4}{2}=8~,
\end{align}
where $N_C$, $N_F$ and $N_S$ are the number of canonical variables, first- and second-class constraints, respectively.
The discussion about proper boundary conditions requires a deeper analysis which is beyond  the scope  of this paper. In the next section, we employ an alternative approach \cite{Heisenberg:2025fxc, Seiler:1995ne}, based on the Cartan-Kuranishi algorithm, to independently verify the number of DOFs and confirm our previous findings.

\subsection{Cartan- Kuranishi algorithm}

An alternative approach to determining the number of DOFs is based on the Cartan–Kuranishi algorithm, which operates directly on the equations of motion. The goal is to identify all hidden integrability conditions in order to obtain an involutive system of partial differential equations. These integrability conditions are revealed by differentiating the equations of motion (prolongation) and projecting the resulting expressions back onto the original system. The counting of DOFs is then performed through a formal Taylor expansion, by enumerating the number of independent coefficients \cite{Heisenberg:2025fxc}. It is worth noting that, for finite-dimensional systems, the Dirac–Bergmann algorithm can be viewed as a related procedure, as it effectively completes the Hamiltonian equations of motion into an involutive system \cite{Seiler:1995ne}. However, naively performing the Dirac-Bergmann algorithm in field theories may not lead to an involutive system of PDEs (see \cite{Seiler:1995ne} for a detailed discussion). In the following, we will perform the Cartan-Kuranishi algorithm to obtain an involutive system  of equations of motion for Eq. \eqref{eq:GR_Parity_violation}.

In the coincident gauge the metric field equations for Eq. \eqref{eq:GR_Parity_violation} are given by
\begin{align}
\label{eq10}
    {\mathcal R_2} & := G_{\mu\nu}+ L_{\mu\nu} = 0~,
\end{align}
where $G_{\mu\nu}$ is the Einstein tensor with respect to the Levi-Civita connection and  $L_{\mu\nu}$ includes the parity-violating terms
\begin{align}
    L_{\mu\nu}=  c_{\rm PV}\epsilon^{\alpha\beta\rho\sigma}\big[\partial_\alpha g_{\beta\nu} \partial_\rho g_{\sigma\mu}- \tensor{g}{^\kappa^\lambda} \partial_\rho \tensor{g}{_\sigma_\kappa}\left(g_{\beta\mu}\partial_\alpha g_{\lambda\nu}+g_{\beta\nu}\partial_\alpha g_{\lambda\mu}\right)\big] ~.
\end{align}
Note that, due to the symmetries of the Levi-Civita tensor, $L_{\mu\nu}$ does not contain second order derivatives of the metric. Therefore, the second order derivative terms or, equivalently, the kinetic matrix are of the same form as in standard GR. 

Before performing the Cartan-Kuranishi algorithm,  we may note that at a point $p$ it is always possible to perform a linear coordinate transformation, fixing $g_{\mu\nu}(p)=\eta_{\mu\nu}$. This redefinition does not spoil the coincident gauge.
However, it is important to emphasize that $\partial_\nu g_{\alpha\beta}(p) \neq 0$, since the adoption of Riemann normal coordinates would be inconsistent with the coincident gauge (see \cite{Heisenberg:2025fxc}  for more details).  

The symbol $S_2$ associated with ${\cal R}_2$ is the same as in GR, since $L_{\mu\nu}$ does not contain second order derivatives. Therefore, we obtain for the $\beta_2^{(k)}$ coefficients
\begin{align}
\beta_2^{(1)}=0\,\,,\,\,\beta_2^{(2)}=0\,\,,\,\,\beta_2^{(3)}=4\,\,,\,\,\beta_2^{(4)}=6~.
\end{align}
Upon prolonging ${\cal R}_3 =\partial {\cal R}_2$, we note that the only contributions involving third derivatives of the metric arise from $\partial_\gamma G_{\mu\nu}$ and it is straightforward to verify that ${\rm rank}S_3 =36$, so that the symbol is involutive, \textit{i.e.} $\sum_{k=1}^4 k\beta_2^{(k)}=36= {\rm rank}\mathcal S_3$, consistent with \cite{Heisenberg:2025fxc}. 

Checking for integrability conditions, we get
\begin{align}
{\rm dim}{\mathcal R_3}-{\rm dim}{\mathcal S_3}\,\,(=136)<{\rm dim}{\mathcal R_2}\,\,(=140)~.
\end{align}
Thus, there are 4 such integrability conditions, which are the well known Bianchi identities, prompting us to perform a prolongation with a subsequent projection. In order to simplify the projection, we can use the Bianchi identity of the Einstein tensor ${\hat \nabla}_\mu G^{\mu\nu}=0$, where ${\hat \nabla}$ is the Levi-Civita covariant derivative. Therefore, we obtain for the 4 independent projected equations
\begin{align}
    g^{\mu\gamma} \partial_\gamma \tensor{G}{_\mu_\nu} + \partial_\gamma L_{\mu\nu} g^{\gamma\mu} =  \left\{\substack{\lambda \\ \beta\alpha}\right\} g^{\beta\alpha} \tensor{G}{_\lambda_\nu} + \left\{\substack{\lambda \\ \gamma\nu}\right\} \tensor{G}{^\gamma_\lambda} + \partial_\gamma L_{\mu\nu} g^{\mu\gamma} =0~,
\end{align}
where $\left\{\substack{\alpha \\ \mu\nu}\right\}$ denotes the Levi-Civita connection.
Finally, the projected system is given by
\begin{align}
    {\cal R}_2^{(1)}: 
    \begin{cases}
        G_{\mu\nu} + L_{\mu\nu} = 0 \\
         \left\{\substack{\lambda \\ \beta\alpha}\right\} g^{\beta\alpha} \tensor{G}{_\lambda_\nu} + \left\{\substack{\lambda \\ \gamma\nu}\right\} \tensor{G}{^\gamma_\lambda} + \partial_\gamma L_{\mu\nu} g^{\mu\gamma} = 0
    \end{cases}
\end{align}
Let us stress again that the system ${\cal R}_2^{(1)}$ is equivalent to ${\cal R}_2$ but it explicitly contains hidden integrability conditions. This is analogous to the Hamiltonian analysis, where first- and second-class constraints are identified. In this context, the four integrability conditions play a role equivalent to that of the four constraints $\{\pi_N, C_r\}$ in the Hamiltonian formulation (see also \cite{Seiler:1995ne} for a comparison  between both approaches). 
To obtain the symbol for ${\cal R}_2^{(1)}$ it is useful to simplify the integrability condition by using $G_{\mu\nu}=-L_{\mu\nu}$, as $L_{\mu\nu}$ only contains first order derivatives. 

Finally, the $\beta$ coefficients of the symbol $S_2^{(1)}$ are given by
\begin{align}
\beta_2^{(1)}=0\,\,,\,\,\beta_2^{(2)}=0\,\,,\,\,\beta_2^{(3)}=4\,\,,\,\,\beta_2^{(4)}=10~.
\end{align}
Similarly, by taking a prolongation of ${\cal R}_2^{(1)}$ it is straightforward to check that the symbol $S_{2}^{(1)}$ is involutive and the system does not contain additional integrability conditions.

As a consistency check, the Cartan characters read
\begin{align}
&\alpha_2^{(1)}=40\,\,,\,\,\alpha_2^{(2)}=30\,\,,\,\,\alpha_2^{(3)}=16\,\,,\,\,\alpha_2^{(4)}=0~.
\end{align}
Given the absence of any residual gauge symmetry, in agreement with $\alpha_2^{(4)}=0$, the number of DOFs is given by
\begin{align}
    \#{\rm DOFs}=m-\frac{1}{q}\beta_2^{(3)}= 10-2=8~,
\end{align}
where $m=10$ is the number of independent fields and $q=2$ is the order of the highest derivatives of ${\cal R}_2^{(1)}$. 
Thus, the theory propagates eight DOFs, consistent with our previous  estimate obtained via the Dirac-Bergmann algorithm.

\section{Conclusions}

In this work, we revisited and extended the framework of Newer General Relativity by incorporating the distinctive quadratic parity-violating operator. We performed a systematic analysis of linear perturbations around a flat FLRW background and demonstrated that the theory generically suffers from ghost instabilities and/ or propagates more DOFs than in the Minkowski limit, thereby already signalling strong coupling. By imposing the requirement of linear stability, we find that only two consistent scenarios arise: STEGR and TDiff-invariant gravity, each supplemented by the parity-violating term. Due to the high symmetry of the background, the parity-violating contribution does not influence the scalar or vector sectors at linear order. Consequently, in STEGR the theory propagates only the two standard tensor DOFs in vacuum, whereas in TDiff-invariant gravity an additional scalar DOF appears in the gravitational sector alongside the two usual tensor modes.

Beyond the linear regime, however, the parity-violating operator explicitly breaks the non-linear TDiff symmetry, inevitably activating further DOFs. Focusing on the framework of STEGR augmented by the parity-violating term, we applied the Dirac-Bergmann algorithm to carry out the Hamiltonian analysis and directly compared its outcomes with those obtained using the recently proposed Cartan-Kuranishi-based method \cite{Heisenberg:2025fxc}. Both approaches consistently show that the theory propagates a total of eight DOFs, corresponding to the six components of the spatial metric together with the two components of the shift vector. This demonstrates that the model enters a strong-coupling regime, restricting the validity of perturbation theory around FLRW. Although this covariant formulation remains a valuable arena for modeling and testing parity-violating effects in gravity, our results indicate that it must be regarded with caution as a fundamental theory, since the emergence of strong coupling is an intrinsic feature of constructions of this type. 

Similar to \cite{DAmbrosio:2023asf,Li:2011rn,Blagojevic:2020dyq,Tomonari:2023wcs} in the case of  $f({\mathbb Q})$ and $f({\mathbb T})$, the Dirac-Bergmann algorithm leads to PDEs for the Lagrange multiplier, breaking the implicit assumption of purely algebraic conditions. As pointed out in \cite{Seiler:1995ne}, due the purely algebraic treatment of the constraints and the fact that prolongations are performed only with respect to the time variable in the Dirac-Bergmann algorithm, one can miss additional constraints/integrability conditions indicating a breakdown of the approach and requiring alternative or extended methods to analyze the constraints. 

Interestingly, a naive application of the Hamiltonian analysis to Eq. \eqref{eq:GR_Parity_violation} yields the same result as the Cartan-Kuranishi algorithm, which avoids the aforementioned limitations. The agreement between the Cartan-Kuranishi analysis and the Dirac-Bergmann algorithm can be understood as follows: all four integrability conditions identified in the Cartan-Kuranishi procedure result from prolongations along the time direction, with none arising from prolongations only along spatial variables. Consequently, they correspond exactly to the Hamiltonian constraints obtained by enforcing the time evolution of the primary constraints. It remains an open question whether this feature can be detected directly from the Hamiltonian analysis; a more detailed investigation of this point is beyond the scope of the present work. Looking ahead, it would be worthwhile to explore the applicability of the Cartan-Kuranishi-based approach to other metric-affine gravity theories or to models in which the Dirac-Bergmann algorithm encounters difficulties. In addition, a systematic study of how to reformulate the Hamiltonian analysis itself as an involutive system would provide a valuable step toward strengthening the correspondence between these two methodologies.

\begin{acknowledgments}
We would like to thank Tomi S. Koivisto, Jose Beltr\'an Jim\'enez and Guillem Domen\`ech for the helpful comments and discussions. 
This research is supported by the DFG under the Emmy-Noether program, project number 496592360.
\end{acknowledgments}

\appendix

\section{Scalar sector: supplementary expressions}

\subsection{Perturbed scalar Lagrangian}
\label{scalarLagrangian}

In this appendix, we present, for completeness, the full scalar Lagrangian, with the TDiff constraint originating from the vector sector (\(2c_1+c_2+c_4=0\)) already imposed. 
\begin{align}
{\mathcal L}^S &= \frac{a^3H^2}{2N}\bigg[3(c_1+3c_3)\partial_i\partial^iE\Big(4(\alpha+3\psi)-\partial_j\partial^jE\Big)-4\Big(3(c_1+3c_3)(2\alpha^2+6\alpha\psi+9\psi^2)\notag\\&+4c_1\partial_i\partial^iE\partial_j\partial^jE-4c_1\partial_i\partial_jE\partial^i\partial^jE\Big)\bigg]-\frac{a^3\dot N}{N^2}\bigg[4(c_1-c_3-c_5)\dot\alpha(4\alpha +6\psi-\partial_i\partial^iE)\notag\\&+(2c_3+c_5)\Big(6\dot\psi(2\alpha+6\psi-\partial_i\partial^iE)-2(\alpha+3\psi)\partial_i\partial^i\dot E+\partial_i\partial^iE\partial_j\partial^jE\Big)\bigg]\notag\\&+\frac{a^3\dot N^2}{2N^3}(c_1-c_3-c_5)\bigg[4(2\alpha^2+6\alpha\psi+9\psi^2)+\partial_i\partial^iE\Big(\partial_j\partial^jE-4(\alpha+3\psi)\Big)\bigg]\notag\\& +\frac{2a^3H}{N}\Big(3(2c_3+c_5)\dot\alpha(4\alpha+6\psi-\partial_i\partial^iE)-6(c_1+3c_3)\dot\psi(2\alpha+6\psi-\partial_i\partial^iE) \notag\\& +(c_1+3c_3)\partial_i\partial^iE(2\alpha+6\psi-\partial_j\partial^jE)\Big)+a^2H\Big(2(4c_1+2c_4-3c_5)(\alpha+3\psi)\partial_i\partial^i\beta \notag\\&+2(4c_1-12c_3+2c_4-7c_5)\partial_i\beta\partial^i\alpha+2(12c_1+36c_3+2c_4+3c_5)\partial_i\beta\partial^i\psi\notag\\& -4(3c_3+2c_5)\partial_i\partial^i\beta\partial_j\partial^jE\Big)+\frac{3a^3H\dot N}{2N^2}(2c_3+c_5)\bigg[4(2\alpha^2+6\alpha\psi+9\psi^2)+\partial_i\partial^iE\notag\\& \Big(\partial_j\partial^jE-4(\alpha+3\psi)\Big)\bigg]-a^2\Big(2(2c_4+c_5)\dot\alpha\partial_i\partial^i\beta+2(4c_1+2c_4-3c_5)\dot\psi\partial_i\partial^i\beta \notag\\& +2(4c_1+2c_4-c_5)\dot\beta\partial_i\partial^i\alpha-2(2c_4+3c_5)\dot\beta\partial_i\partial^i\psi-(4c_1+2c_4-c_5)\partial_i\partial^i\beta\partial_j\partial^j\dot E \notag\\& -(2c_4+c_5)\partial_i\partial^i\dot\beta\partial_j\partial^jE\Big)+\frac{a^2\dot N}{N}\Big(2(2c_4+c_5)(\alpha+3\psi)\partial_i\partial^i\beta +2(4c_3+2c_4+3c_5)\partial_i\beta\partial^i\alpha \notag\\& -2(12c_3-2c_4+3c_5)\partial_i\beta\partial^i\psi-2(2c_3+c_5)\partial_i\partial^i\beta\partial_j\partial^jE\Big) +\frac{a^3}{N}\bigg[4(c_{1}-c_{3}-c_{5})\dot{\alpha}^{2} \notag\\& -(c_{1}+c_{3})\partial_i \partial^i\dot{E}\partial_j \partial^j\dot{E} -12(c_{1}+3 c_{3})\dot{\psi}^{2}+2(2 c_{3}+c_{5})\dot{\alpha}  (6\dot\psi-\partial_i \partial^i\dot{E}) +4(c_{1}+3 c_{3})\partial_i \partial^i\dot{E}\dot{\psi}\bigg]\notag\\&  -Na\bigg[-4(c_1+c_3)\partial_i\alpha\partial^i\alpha+4\Big((6c_3+c_5)\partial^i\alpha -(c_1+9c_3+3c_5)\partial^i\psi\Big)\partial_i\psi\notag \\& -2\Big((2c_3+c_5)\partial^i\alpha+2(c_1-3c_3-2c_5)\partial^i\psi\Big)\partial_i\partial_j\partial^jE  +(c_1-c_3-c_5)\partial_i\partial_j\partial_kE\partial^i\partial^j\partial^kE\bigg]
\end{align}

\newpage

\subsection{Scalar Lagrangian: kinetic terms}
\label{subsec:Kinetic_terms}
 In this appendix, we present the kinetic terms $K_{ij}$ appearing in Eq. \eqref{eq:Lagrangian_Two_scalar}, whose explicit forms are 
 \begin{align}
     K_{11} =& - \frac{1}{16 (c_1 + c_4) ( 4 c_1 (c_1 + 2 c_3) - 4 c_1 c_5 + 3 c_5^2 ) H^2 N^2   } \Big[ (8 c_1 (c_1 (5 c_1^2 + 41 c_1 c_3 + 74 c_3^2) + 
    4 (c_1 + c_3) (c_1 + 2 c_3) c_4) \nonumber \\
    & + 
 4 c_1 (7 c_1^2 - 31 c_1 c_3 - 84 c_3^2 - 8 (c_1 + c_3) c_4) c_5 + 
 2 (-25 c_1^2 + 51 c_1 c_3 + 18 c_3^2 + 12 (c_1 + c_3) c_4) c_5^2 \nonumber \\
 & + 
 3 (19 c_1 - 15 c_3) c_5^3 - 18 c_5^4 ) H^2 N^2 - ((2 c_1 - 4 c_3 - 3 c_5) (2 c_1 - c_5) (4 c_1 (2 c_1 + 7 c_3) \nonumber \\
 & - 2 (c_1 + 3 c_3) c_5 + 3  c_5^2)) H N \dot N +(2 c_1 - 4 c_3 - 3 c_5)  (c_1 - c_3 - c_5)  (-2 c_1 + c_5)^2 \dot N^2  \Big] \\
 K_{12} = & \frac{2 c_1 }{(c_1 + c_4) (4 c_1 (c_1 + 2 c_3) - 4 c_1 c_5 + 3  c_5^2) H^2 N^2} \Big[ ( 4 c_1 (3 c_1^2 + 15 c_1 c_2 + 30 c_2^2 + 2 c_1 c_4 + 4 c_2 c_4) \nonumber \\ & + 
 36 (c_1 - c_2) c_2 c_5  - 8 c_1 c_4 c_5 + 3 (3 c_1 - 3 c_2 + 2 c_4) c_5^2 ) H^2 N^2 + ( -8 c_1^3 + 4 c_1^2 (-8 c_2 + c_5) - 3 c_5 \nonumber \\
& \times (8 c_3^2 + 8 c_3 c_5 + c_5^2) + 
 2 c_1 (32 c_3^2 + 40 c_3 c_5 + 5 c_5^2) ) H N \dot N + (2 c_1 - 4 c_3 - 3 c_5) (2 c_1 - c_5) (c_1 - c_3 - c_5) \dot N^2 \Big] \\
 K_{22} =& \frac{4 c_1 }{(c_1 + c_4) (4 c_1 (c_1 + 2 c_3) - 4 c_1 c_5 + 3  c_5^2) H^2 N^2 } \Big[ (-3 c_1 (2 c_1^2 + 6 c_1 c_3 + 12 c_3^2 - c_1 c_5 + 9 c_3 c_5 + 3  c_5^2)) H^2 N^2 \nonumber \\
 & +(12 c_1 (c_1 - 2 c_3) c_3 - 4 c_1 (c_1 + 2 c_3) c_4 + 
   2 c_1 (3 c_1 - 15 c_3 + 2 c_4) c_5 - 3  (3 c_1 + c_4)  c_5^2) H N \dot N \nonumber \\
   & -c_1  (2 c_1 - 4 c_3 - 3 c_5)  (c_1 - c_3 - c_5) \dot N^2\Big]~.
 \end{align}
 To determine the conditions for a degenerate kinetic matrix, we require that the degeneracy holds for an arbitrary background evolution. This leads to four independent conditions that must be satisfied simultaneously.

\section{Uniqueness of solutions for the Lagrange multipliers}
\label{app:Lagrange_mulitplier}

In this appendix, we provide a more detailed structural analysis of the PDEs \eqref{eq:Lagrange_Multiplier_1} and \eqref{eq:Lagrange_Multiplier_2}, focusing on their solvability properties.
Let us start from decomposing \(\vec u\) into a component parallel to \(\vec \pi\) and two components orthogonal to \(\vec \pi\). Thus, we choose an orthonormal triad \(\left\{e_r^{(1)}, e_r^{(2)}, e_r^{(3)}\right\}\) with \(e_r^{(1)}\equiv\pi_r/| \pi_r|\), where $\vert \pi_r \vert = \sqrt{\gamma_{ij} \pi^i\pi^j}$, getting
\begin{align}
    u_r=u_{\pi}\frac{\pi_r}{| \pi_r|^2}+\sum_{A=2,3}u_{\perp}^{(A)}e^{(A)}_r~,
\end{align}
where $u_{\pi}:=u^r\pi_r$ and, of course, $e^{{(A)}}_r \pi^r=0$ with  $A=2,3$.
At this point, it is immediate to see that Eq. \eqref{eq:Lagrange_Multiplier_1} is an algebraic equation fixing \(u_{\pi}\)
\begin{align}
    u_{\pi}\approx \frac{N}{3}\{\pi_N,{\cal H}_0\}~.
\end{align}
Next, we project Eq. \eqref{eq:Lagrange_Multiplier_2} along $e^{(1)r} $, which can be solved for $u_N$, leading to
\begin{align}
    \frac{3}{N} u_N \vert \pi_r \vert \approx - e^{(1)r} \{ C_r. {\cal H}_0\} - 4 c_{\rm PV} \sqrt{\gamma} \tensor{\epsilon}{_l^j_r} e^{(1)r} {\cal D}_j \left( \frac{u^l}{N^3} \right)~.
\end{align}
Finally, the two  non-algebraic equations for $u^{(A)}_\perp$ are of the form
\begin{align}
    \frac{4 c_{\rm PV} \sqrt{\gamma}}{N^3 }\tensor{\epsilon}{_l^j_r} e^{(A)r} {\cal D}_j \left( \sum_{B=2,3} u_\perp^{(B)} e^{(B)l} \right) \approx  e^{(A)r} B_r(u_\pi, u_\perp )~,
\end{align}
where $B_r$ contains all the terms which do not depend on derivatives of $u_\perp^{(A)}$ and whose precise form is not relevant for our purpose. Note that the projected equations along  $e^{(A)r}$ do not depend on $u_N$ as $\pi_r e^{(A)r}=0$.

The two equations are linear order PDEs which are of the general form
\begin{align}
    \sum_k^3 M^{(k)} \partial_k u^{(A)}_\perp + L u^{(A)}_\perp + V \approx 0~.
\end{align}
In our case the matrix  \(M\) is given by
\begin{align}
    M^{(j)}_{AB}:=\frac{4 c_{\rm PV}\sqrt{\gamma}}{N^3}\tensor{\epsilon}{_l^j_r} e^{(A)r} e^{(B)l}~,
\end{align}
and the form of  $L$ and $V$ is again not relevant for  our purpose. 
Following \cite{DAmbrosio:2023asf,Heisenberg:2025fxc}, the PDEs have a unique solution for $u_\perp^{(A)}$ if and only if  the matrix 
\begin{align}
    \sigma(s)_{AB}:=\sum_{j=1}^3M^{(j)}_{AB}s_j = \frac{4 c_{\rm PV}\sqrt{\gamma}}{N^3}(\tensor{\epsilon}{_l^j_r}s_j) e^{(A)r} e^{(B)l}
\end{align}
is invertible,  where $s_k$ is the normal vector to the characteristic surface on which the boundary conditions for $u_\perp^{(A)}$ are defined. 

Recalling that \(\left(e^{(A)}\cross e^{(B)}\right)^j=\tensor{\epsilon}{^j_l_r} e^{(A)l} e^{(B)r}\) and, due to the orthonormality of the basis, \(\left(e^{(2)}\cross e^{(3)}\right)^j=e^{(1)j}\), we have
\begin{align}
    \sigma(s) = \frac{4 c_{\rm PV}\sqrt{\gamma}}{N^3}\frac{ \pi^j s_j}{|\pi_r|} \begin{pmatrix}0 & 1 \\-1 & 0 \end{pmatrix}~.
\end{align}
Thus
\begin{align}
    {\rm det}\left(\sigma(s)\right) = \left(\frac{4 c_{\rm PV}\sqrt{\gamma}}{N^3| \pi_r|}\right)^2\left( \pi^j s_j\right)^2\not \approx 0 \iff \vec \pi\cdot\vec s\not \approx 0~.
\end{align}
The condition on the characteristic surface for the uniqueness of the solution  can be easily understood by noting that, if the characteristic plane includes  the vector $\pi_r$, the algebraic condition and the boundary conditions are all lying in the same plane, so that the derivative orthogonal to such plane remains unconstrained.

It is noteworthy that in the Hamiltonian analysis of $f({\mathbb T})$ gravity carried out in \cite{Li:2011rn,Blagojevic:2020dyq,Tomonari:2023wcs}, the authors obtained a PDE for the final Lagrange multiplier of the form
\begin{align}
    A^k\partial_k u + L u +  G \approx 0~.
\end{align}
Therefore, under the assumption $A^k \not  \approx  0$, the PDE admits a unique solution if and only if the normal to the characteristic surface $s^j$ is not orthogonal to $A^k$, \textit{i.e.} $A^k s_k \not \approx 0$.

\bibliography{refs.bib}

\begin{thebibliography}{59}%
\makeatletter
\providecommand \@ifxundefined [1]{%
 \@ifx{#1\undefined}
}%
\providecommand \@ifnum [1]{%
 \ifnum #1\expandafter \@firstoftwo
 \else \expandafter \@secondoftwo
 \fi
}%
\providecommand \@ifx [1]{%
 \ifx #1\expandafter \@firstoftwo
 \else \expandafter \@secondoftwo
 \fi
}%
\providecommand \natexlab [1]{#1}%
\providecommand \enquote  [1]{``#1''}%
\providecommand \bibnamefont  [1]{#1}%
\providecommand \bibfnamefont [1]{#1}%
\providecommand \citenamefont [1]{#1}%
\providecommand \href@noop [0]{\@secondoftwo}%
\providecommand \href [0]{\begingroup \@sanitize@url \@href}%
\providecommand \@href[1]{\@@startlink{#1}\@@href}%
\providecommand \@@href[1]{\endgroup#1\@@endlink}%
\providecommand \@sanitize@url [0]{\catcode `\\12\catcode `\$12\catcode
  `\&12\catcode `\#12\catcode `\^12\catcode `\_12\catcode `\%12\relax}%
\providecommand \@@startlink[1]{}%
\providecommand \@@endlink[0]{}%
\providecommand \url  [0]{\begingroup\@sanitize@url \@url }%
\providecommand \@url [1]{\endgroup\@href {#1}{\urlprefix }}%
\providecommand \urlprefix  [0]{URL }%
\providecommand \Eprint [0]{\href }%
\providecommand \doibase [0]{https://doi.org/}%
\providecommand \selectlanguage [0]{\@gobble}%
\providecommand \bibinfo  [0]{\@secondoftwo}%
\providecommand \bibfield  [0]{\@secondoftwo}%
\providecommand \translation [1]{[#1]}%
\providecommand \BibitemOpen [0]{}%
\providecommand \bibitemStop [0]{}%
\providecommand \bibitemNoStop [0]{.\EOS\space}%
\providecommand \EOS [0]{\spacefactor3000\relax}%
\providecommand \BibitemShut  [1]{\csname bibitem#1\endcsname}%
\let\auto@bib@innerbib\@empty
\bibitem [{\citenamefont {Élie Cartan}(1922)}]{Cartan1922}%
  \BibitemOpen
  \bibfield  {author} {\bibinfo {author} {\bibnamefont {Élie Cartan}},\
  }\href@noop {} {\bibfield  {journal} {\bibinfo  {journal} {Comptes Rendus de
  l’Académie des Sciences}\ }\textbf {\bibinfo {volume} {174}},\ \bibinfo
  {pages} {593} (\bibinfo {year} {1922})}\BibitemShut {NoStop}%
\bibitem [{\citenamefont {Élie Cartan}(1923)}]{Cartan1923}%
  \BibitemOpen
  \bibfield  {author} {\bibinfo {author} {\bibnamefont {Élie Cartan}},\
  }\href@noop {} {\bibfield  {journal} {\bibinfo  {journal} {Annales
  Scientifiques de l'École Normale Supérieure}\ }\textbf {\bibinfo {volume}
  {40}},\ \bibinfo {pages} {325} (\bibinfo {year} {1923})}\BibitemShut
  {NoStop}%
\bibitem [{\citenamefont {Élie Cartan}(1924)}]{Cartan1924}%
  \BibitemOpen
  \bibfield  {author} {\bibinfo {author} {\bibnamefont {Élie Cartan}},\
  }\href@noop {} {\bibfield  {journal} {\bibinfo  {journal} {Annales
  Scientifiques de l'École Normale Supérieure}\ }\textbf {\bibinfo {volume}
  {41}},\ \bibinfo {pages} {1} (\bibinfo {year} {1924})}\BibitemShut {NoStop}%
\bibitem [{\citenamefont {Élie Cartan}(1925)}]{Cartan1925}%
  \BibitemOpen
  \bibfield  {author} {\bibinfo {author} {\bibnamefont {Élie Cartan}},\
  }\href@noop {} {\bibfield  {journal} {\bibinfo  {journal} {Annales
  Scientifiques de l'École Normale Supérieure}\ }\textbf {\bibinfo {volume}
  {42}},\ \bibinfo {pages} {17} (\bibinfo {year} {1925})}\BibitemShut {NoStop}%
\bibitem [{\citenamefont {Hehl}\ \emph {et~al.}(1976)\citenamefont {Hehl},
  \citenamefont {Von Der~Heyde}, \citenamefont {Kerlick},\ and\ \citenamefont
  {Nester}}]{Hehl:1976kj}%
  \BibitemOpen
  \bibfield  {author} {\bibinfo {author} {\bibfnamefont {F.~W.}\ \bibnamefont
  {Hehl}}, \bibinfo {author} {\bibfnamefont {P.}~\bibnamefont {Von Der~Heyde}},
  \bibinfo {author} {\bibfnamefont {G.~D.}\ \bibnamefont {Kerlick}},\ and\
  \bibinfo {author} {\bibfnamefont {J.~M.}\ \bibnamefont {Nester}},\ }\href
  {https://doi.org/10.1103/RevModPhys.48.393} {\bibfield  {journal} {\bibinfo
  {journal} {Rev. Mod. Phys.}\ }\textbf {\bibinfo {volume} {48}},\ \bibinfo
  {pages} {393} (\bibinfo {year} {1976})}\BibitemShut {NoStop}%
\bibitem [{\citenamefont {Ferraris}\ and\ \citenamefont
  {Kijowski}(1982)}]{Ferraris:1982ez}%
  \BibitemOpen
  \bibfield  {author} {\bibinfo {author} {\bibfnamefont {M.}~\bibnamefont
  {Ferraris}}\ and\ \bibinfo {author} {\bibfnamefont {J.}~\bibnamefont
  {Kijowski}},\ }\href {https://doi.org/10.1007/BF00756921} {\bibfield
  {journal} {\bibinfo  {journal} {Gen. Rel. Grav.}\ }\textbf {\bibinfo {volume}
  {14}},\ \bibinfo {pages} {165} (\bibinfo {year} {1982})}\BibitemShut
  {NoStop}%
\bibitem [{\citenamefont {Nester}\ and\ \citenamefont
  {Yo}(1999)}]{Nester:1998mp}%
  \BibitemOpen
  \bibfield  {author} {\bibinfo {author} {\bibfnamefont {J.~M.}\ \bibnamefont
  {Nester}}\ and\ \bibinfo {author} {\bibfnamefont {H.-J.}\ \bibnamefont
  {Yo}},\ }\href@noop {} {\bibfield  {journal} {\bibinfo  {journal} {Chin. J.
  Phys.}\ }\textbf {\bibinfo {volume} {37}},\ \bibinfo {pages} {113} (\bibinfo
  {year} {1999})},\ \Eprint {https://arxiv.org/abs/gr-qc/9809049}
  {arXiv:gr-qc/9809049} \BibitemShut {NoStop}%
\bibitem [{\citenamefont {Beltr{\'a}n~Jim{\'e}nez}\ \emph
  {et~al.}(2018{\natexlab{a}})\citenamefont {Beltr{\'a}n~Jim{\'e}nez},
  \citenamefont {Heisenberg},\ and\ \citenamefont
  {Koivisto}}]{BeltranJimenez:2017tkd}%
  \BibitemOpen
  \bibfield  {author} {\bibinfo {author} {\bibfnamefont {J.}~\bibnamefont
  {Beltr{\'a}n~Jim{\'e}nez}}, \bibinfo {author} {\bibfnamefont
  {L.}~\bibnamefont {Heisenberg}},\ and\ \bibinfo {author} {\bibfnamefont
  {T.}~\bibnamefont {Koivisto}},\ }\href
  {https://doi.org/10.1103/PhysRevD.98.044048} {\bibfield  {journal} {\bibinfo
  {journal} {Phys. Rev. D}\ }\textbf {\bibinfo {volume} {98}},\ \bibinfo
  {pages} {044048} (\bibinfo {year} {2018}{\natexlab{a}})},\ \Eprint
  {https://arxiv.org/abs/1710.03116} {arXiv:1710.03116 [gr-qc]} \BibitemShut
  {NoStop}%
\bibitem [{\citenamefont {Beltr{\'a}n~Jim{\'e}nez}\ \emph
  {et~al.}(2019)\citenamefont {Beltr{\'a}n~Jim{\'e}nez}, \citenamefont
  {Heisenberg},\ and\ \citenamefont {Koivisto}}]{BeltranJimenez:2019esp}%
  \BibitemOpen
  \bibfield  {author} {\bibinfo {author} {\bibfnamefont {J.}~\bibnamefont
  {Beltr{\'a}n~Jim{\'e}nez}}, \bibinfo {author} {\bibfnamefont
  {L.}~\bibnamefont {Heisenberg}},\ and\ \bibinfo {author} {\bibfnamefont
  {T.~S.}\ \bibnamefont {Koivisto}},\ }\href
  {https://doi.org/10.3390/universe5070173} {\bibfield  {journal} {\bibinfo
  {journal} {Universe}\ }\textbf {\bibinfo {volume} {5}},\ \bibinfo {pages}
  {173} (\bibinfo {year} {2019})},\ \Eprint {https://arxiv.org/abs/1903.06830}
  {arXiv:1903.06830 [hep-th]} \BibitemShut {NoStop}%
\bibitem [{\citenamefont {de~Andrade}\ and\ \citenamefont
  {Pereira}(1999)}]{deAndrade:1997cj}%
  \BibitemOpen
  \bibfield  {author} {\bibinfo {author} {\bibfnamefont {V.~C.}\ \bibnamefont
  {de~Andrade}}\ and\ \bibinfo {author} {\bibfnamefont {J.~G.}\ \bibnamefont
  {Pereira}},\ }\href {https://doi.org/10.1142/S0218271899000122} {\bibfield
  {journal} {\bibinfo  {journal} {Int. J. Mod. Phys. D}\ }\textbf {\bibinfo
  {volume} {8}},\ \bibinfo {pages} {141} (\bibinfo {year} {1999})},\ \Eprint
  {https://arxiv.org/abs/gr-qc/9708051} {arXiv:gr-qc/9708051} \BibitemShut
  {NoStop}%
\bibitem [{\citenamefont {Beltr{\'a}n~Jim{\'e}nez}\ \emph
  {et~al.}(2020)\citenamefont {Beltr{\'a}n~Jim{\'e}nez}, \citenamefont
  {Heisenberg},\ and\ \citenamefont {Koivisto}}]{BeltranJimenez:2020sih}%
  \BibitemOpen
  \bibfield  {author} {\bibinfo {author} {\bibfnamefont {J.}~\bibnamefont
  {Beltr{\'a}n~Jim{\'e}nez}}, \bibinfo {author} {\bibfnamefont
  {L.}~\bibnamefont {Heisenberg}},\ and\ \bibinfo {author} {\bibfnamefont
  {T.}~\bibnamefont {Koivisto}},\ }\href
  {https://doi.org/10.1088/1361-6382/aba31b} {\bibfield  {journal} {\bibinfo
  {journal} {Class. Quant. Grav.}\ }\textbf {\bibinfo {volume} {37}},\ \bibinfo
  {pages} {195013} (\bibinfo {year} {2020})},\ \Eprint
  {https://arxiv.org/abs/2004.04606} {arXiv:2004.04606 [hep-th]} \BibitemShut
  {NoStop}%
\bibitem [{\citenamefont {So}\ and\ \citenamefont {Nester}(2006)}]{So:2006pm}%
  \BibitemOpen
  \bibfield  {author} {\bibinfo {author} {\bibfnamefont {L.~L.}\ \bibnamefont
  {So}}\ and\ \bibinfo {author} {\bibfnamefont {J.~M.}\ \bibnamefont
  {Nester}},\ }in\ \href@noop {} {\emph {\bibinfo {booktitle} {{10th Marcel
  Grossmann Meeting on Recent Developments in Theoretical and Experimental
  General Relativity, Gravitation and Relativistic Field Theories (MG X
  MMIII)}}}}\ (\bibinfo {year} {2006})\ \Eprint
  {https://arxiv.org/abs/gr-qc/0612062} {arXiv:gr-qc/0612062} \BibitemShut
  {NoStop}%
\bibitem [{\citenamefont {Delhom}(2020)}]{Delhom:2020hkb}%
  \BibitemOpen
  \bibfield  {author} {\bibinfo {author} {\bibfnamefont {A.}~\bibnamefont
  {Delhom}},\ }\href {https://doi.org/10.1140/epjc/s10052-020-8330-y}
  {\bibfield  {journal} {\bibinfo  {journal} {Eur. Phys. J. C}\ }\textbf
  {\bibinfo {volume} {80}},\ \bibinfo {pages} {728} (\bibinfo {year} {2020})},\
  \Eprint {https://arxiv.org/abs/2002.02404} {arXiv:2002.02404 [gr-qc]}
  \BibitemShut {NoStop}%
\bibitem [{\citenamefont {Heisenberg}(2024)}]{Heisenberg:2023lru}%
  \BibitemOpen
  \bibfield  {author} {\bibinfo {author} {\bibfnamefont {L.}~\bibnamefont
  {Heisenberg}},\ }\href {https://doi.org/10.1016/j.physrep.2024.02.001}
  {\bibfield  {journal} {\bibinfo  {journal} {Phys. Rept.}\ }\textbf {\bibinfo
  {volume} {1066}},\ \bibinfo {pages} {1} (\bibinfo {year} {2024})},\ \Eprint
  {https://arxiv.org/abs/2309.15958} {arXiv:2309.15958 [gr-qc]} \BibitemShut
  {NoStop}%
\bibitem [{\citenamefont {Cai}\ \emph {et~al.}(2016)\citenamefont {Cai},
  \citenamefont {Capozziello}, \citenamefont {De~Laurentis},\ and\
  \citenamefont {Saridakis}}]{Cai:2015emx}%
  \BibitemOpen
  \bibfield  {author} {\bibinfo {author} {\bibfnamefont {Y.-F.}\ \bibnamefont
  {Cai}}, \bibinfo {author} {\bibfnamefont {S.}~\bibnamefont {Capozziello}},
  \bibinfo {author} {\bibfnamefont {M.}~\bibnamefont {De~Laurentis}},\ and\
  \bibinfo {author} {\bibfnamefont {E.~N.}\ \bibnamefont {Saridakis}},\ }\href
  {https://doi.org/10.1088/0034-4885/79/10/106901} {\bibfield  {journal}
  {\bibinfo  {journal} {Rept. Prog. Phys.}\ }\textbf {\bibinfo {volume} {79}},\
  \bibinfo {pages} {106901} (\bibinfo {year} {2016})},\ \Eprint
  {https://arxiv.org/abs/1511.07586} {arXiv:1511.07586 [gr-qc]} \BibitemShut
  {NoStop}%
\bibitem [{\citenamefont {Hayashi}\ and\ \citenamefont
  {Shirafuji}(1979)}]{PhysRevD.19.3524}%
  \BibitemOpen
  \bibfield  {author} {\bibinfo {author} {\bibfnamefont {K.}~\bibnamefont
  {Hayashi}}\ and\ \bibinfo {author} {\bibfnamefont {T.}~\bibnamefont
  {Shirafuji}},\ }\href {https://doi.org/10.1103/PhysRevD.19.3524} {\bibfield
  {journal} {\bibinfo  {journal} {Phys. Rev. D}\ }\textbf {\bibinfo {volume}
  {19}},\ \bibinfo {pages} {3524} (\bibinfo {year} {1979})}\BibitemShut
  {NoStop}%
\bibitem [{\citenamefont {Iosifidis}\ and\ \citenamefont
  {Koivisto}(2019)}]{Iosifidis:2018zwo}%
  \BibitemOpen
  \bibfield  {author} {\bibinfo {author} {\bibfnamefont {D.}~\bibnamefont
  {Iosifidis}}\ and\ \bibinfo {author} {\bibfnamefont {T.}~\bibnamefont
  {Koivisto}},\ }\href {https://doi.org/10.3390/universe5030082} {\bibfield
  {journal} {\bibinfo  {journal} {Universe}\ }\textbf {\bibinfo {volume} {5}},\
  \bibinfo {pages} {82} (\bibinfo {year} {2019})},\ \Eprint
  {https://arxiv.org/abs/1810.12276} {arXiv:1810.12276 [gr-qc]} \BibitemShut
  {NoStop}%
\bibitem [{\citenamefont {Conroy}\ and\ \citenamefont
  {Koivisto}(2019)}]{Conroy:2019ibo}%
  \BibitemOpen
  \bibfield  {author} {\bibinfo {author} {\bibfnamefont {A.}~\bibnamefont
  {Conroy}}\ and\ \bibinfo {author} {\bibfnamefont {T.}~\bibnamefont
  {Koivisto}},\ }\href {https://doi.org/10.1088/1475-7516/2019/12/016}
  {\bibfield  {journal} {\bibinfo  {journal} {JCAP}\ }\textbf {\bibinfo
  {volume} {12}},\ \bibinfo {pages} {016}},\ \Eprint
  {https://arxiv.org/abs/1908.04313} {arXiv:1908.04313 [gr-qc]} \BibitemShut
  {NoStop}%
\bibitem [{\citenamefont {Chen}\ \emph {et~al.}(2023)\citenamefont {Chen},
  \citenamefont {Yu},\ and\ \citenamefont {Gao}}]{Chen:2022wtz}%
  \BibitemOpen
  \bibfield  {author} {\bibinfo {author} {\bibfnamefont {Z.}~\bibnamefont
  {Chen}}, \bibinfo {author} {\bibfnamefont {Y.}~\bibnamefont {Yu}},\ and\
  \bibinfo {author} {\bibfnamefont {X.}~\bibnamefont {Gao}},\ }\href
  {https://doi.org/10.1088/1475-7516/2023/06/001} {\bibfield  {journal}
  {\bibinfo  {journal} {JCAP}\ }\textbf {\bibinfo {volume} {06}},\ \bibinfo
  {pages} {001}},\ \Eprint {https://arxiv.org/abs/2212.14362} {arXiv:2212.14362
  [gr-qc]} \BibitemShut {NoStop}%
\bibitem [{\citenamefont {Beltr{\'a}n~Jim{\'e}nez}\ \emph
  {et~al.}(2018{\natexlab{b}})\citenamefont {Beltr{\'a}n~Jim{\'e}nez},
  \citenamefont {Heisenberg},\ and\ \citenamefont
  {Koivisto}}]{BeltranJimenez:2018vdo}%
  \BibitemOpen
  \bibfield  {author} {\bibinfo {author} {\bibfnamefont {J.}~\bibnamefont
  {Beltr{\'a}n~Jim{\'e}nez}}, \bibinfo {author} {\bibfnamefont
  {L.}~\bibnamefont {Heisenberg}},\ and\ \bibinfo {author} {\bibfnamefont
  {T.~S.}\ \bibnamefont {Koivisto}},\ }\href
  {https://doi.org/10.1088/1475-7516/2018/08/039} {\bibfield  {journal}
  {\bibinfo  {journal} {JCAP}\ }\textbf {\bibinfo {volume} {08}},\ \bibinfo
  {pages} {039}},\ \Eprint {https://arxiv.org/abs/1803.10185} {arXiv:1803.10185
  [gr-qc]} \BibitemShut {NoStop}%
\bibitem [{\citenamefont {Bello-Morales}\ \emph {et~al.}(2024)\citenamefont
  {Bello-Morales}, \citenamefont {Beltr{\'a}n~Jim{\'e}nez}, \citenamefont
  {Jim{\'e}nez~Cano}, \citenamefont {Koivisto},\ and\ \citenamefont
  {Maroto}}]{Bello-Morales:2024vqk}%
  \BibitemOpen
  \bibfield  {author} {\bibinfo {author} {\bibfnamefont {A.~G.}\ \bibnamefont
  {Bello-Morales}}, \bibinfo {author} {\bibfnamefont {J.}~\bibnamefont
  {Beltr{\'a}n~Jim{\'e}nez}}, \bibinfo {author} {\bibfnamefont
  {A.}~\bibnamefont {Jim{\'e}nez~Cano}}, \bibinfo {author} {\bibfnamefont
  {T.~S.}\ \bibnamefont {Koivisto}},\ and\ \bibinfo {author} {\bibfnamefont
  {A.~L.}\ \bibnamefont {Maroto}},\ }\href
  {https://doi.org/10.1007/JHEP12(2024)146} {\bibfield  {journal} {\bibinfo
  {journal} {JHEP}\ }\textbf {\bibinfo {volume} {12}},\ \bibinfo {pages}
  {146}},\ \Eprint {https://arxiv.org/abs/2406.19355} {arXiv:2406.19355
  [gr-qc]} \BibitemShut {NoStop}%
\bibitem [{\citenamefont {Alvarez}\ \emph {et~al.}(2006)\citenamefont
  {Alvarez}, \citenamefont {Blas}, \citenamefont {Garriga},\ and\ \citenamefont
  {Verdaguer}}]{Alvarez:2006uu}%
  \BibitemOpen
  \bibfield  {author} {\bibinfo {author} {\bibfnamefont {E.}~\bibnamefont
  {Alvarez}}, \bibinfo {author} {\bibfnamefont {D.}~\bibnamefont {Blas}},
  \bibinfo {author} {\bibfnamefont {J.}~\bibnamefont {Garriga}},\ and\ \bibinfo
  {author} {\bibfnamefont {E.}~\bibnamefont {Verdaguer}},\ }\href
  {https://doi.org/10.1016/j.nuclphysb.2006.08.003} {\bibfield  {journal}
  {\bibinfo  {journal} {Nucl. Phys. B}\ }\textbf {\bibinfo {volume} {756}},\
  \bibinfo {pages} {148} (\bibinfo {year} {2006})},\ \Eprint
  {https://arxiv.org/abs/hep-th/0606019} {arXiv:hep-th/0606019} \BibitemShut
  {NoStop}%
\bibitem [{\citenamefont {Alvarez}\ \emph {et~al.}(2009)\citenamefont
  {Alvarez}, \citenamefont {Faedo},\ and\ \citenamefont
  {Lopez-Villarejo}}]{Alvarez:2009ga}%
  \BibitemOpen
  \bibfield  {author} {\bibinfo {author} {\bibfnamefont {E.}~\bibnamefont
  {Alvarez}}, \bibinfo {author} {\bibfnamefont {A.~F.}\ \bibnamefont {Faedo}},\
  and\ \bibinfo {author} {\bibfnamefont {J.~J.}\ \bibnamefont
  {Lopez-Villarejo}},\ }\href {https://doi.org/10.1088/1475-7516/2009/07/002}
  {\bibfield  {journal} {\bibinfo  {journal} {JCAP}\ }\textbf {\bibinfo
  {volume} {07}},\ \bibinfo {pages} {002}},\ \Eprint
  {https://arxiv.org/abs/0904.3298} {arXiv:0904.3298 [hep-th]} \BibitemShut
  {NoStop}%
\bibitem [{\citenamefont {Blas}\ \emph {et~al.}(2011)\citenamefont {Blas},
  \citenamefont {Shaposhnikov},\ and\ \citenamefont
  {Zenhausern}}]{Blas:2011ac}%
  \BibitemOpen
  \bibfield  {author} {\bibinfo {author} {\bibfnamefont {D.}~\bibnamefont
  {Blas}}, \bibinfo {author} {\bibfnamefont {M.}~\bibnamefont {Shaposhnikov}},\
  and\ \bibinfo {author} {\bibfnamefont {D.}~\bibnamefont {Zenhausern}},\
  }\href {https://doi.org/10.1103/PhysRevD.84.044001} {\bibfield  {journal}
  {\bibinfo  {journal} {Phys. Rev. D}\ }\textbf {\bibinfo {volume} {84}},\
  \bibinfo {pages} {044001} (\bibinfo {year} {2011})},\ \Eprint
  {https://arxiv.org/abs/1104.1392} {arXiv:1104.1392 [hep-th]} \BibitemShut
  {NoStop}%
\bibitem [{\citenamefont {Bello-Morales}\ and\ \citenamefont
  {Maroto}(2024)}]{Bello-Morales:2023btf}%
  \BibitemOpen
  \bibfield  {author} {\bibinfo {author} {\bibfnamefont {A.~G.}\ \bibnamefont
  {Bello-Morales}}\ and\ \bibinfo {author} {\bibfnamefont {A.~L.}\ \bibnamefont
  {Maroto}},\ }\href {https://doi.org/10.1103/PhysRevD.109.043506} {\bibfield
  {journal} {\bibinfo  {journal} {Phys. Rev. D}\ }\textbf {\bibinfo {volume}
  {109}},\ \bibinfo {pages} {043506} (\bibinfo {year} {2024})},\ \Eprint
  {https://arxiv.org/abs/2308.00635} {arXiv:2308.00635 [gr-qc]} \BibitemShut
  {NoStop}%
\bibitem [{\citenamefont {Ganz}(2022)}]{Ganz:2022iiv}%
  \BibitemOpen
  \bibfield  {author} {\bibinfo {author} {\bibfnamefont {A.}~\bibnamefont
  {Ganz}},\ }\href {https://doi.org/10.1088/1475-7516/2022/08/074} {\bibfield
  {journal} {\bibinfo  {journal} {JCAP}\ }\textbf {\bibinfo {volume} {08}},\
  \bibinfo {pages} {074}},\ \Eprint {https://arxiv.org/abs/2203.12358}
  {arXiv:2203.12358 [gr-qc]} \BibitemShut {NoStop}%
\bibitem [{\citenamefont {Dirac}(1958)}]{Dirac:1958sq}%
  \BibitemOpen
  \bibfield  {author} {\bibinfo {author} {\bibfnamefont {P.~A.~M.}\
  \bibnamefont {Dirac}},\ }\href {https://doi.org/10.1098/rspa.1958.0141}
  {\bibfield  {journal} {\bibinfo  {journal} {Proc. Roy. Soc. Lond. A}\
  }\textbf {\bibinfo {volume} {246}},\ \bibinfo {pages} {326} (\bibinfo {year}
  {1958})}\BibitemShut {NoStop}%
\bibitem [{\citenamefont {Dirac}(2001)}]{dirac2001lectures}%
  \BibitemOpen
  \bibfield  {author} {\bibinfo {author} {\bibfnamefont {P.~A.~M.}\
  \bibnamefont {Dirac}},\ }\href@noop {} {\emph {\bibinfo {title} {Lectures on
  quantum mechanics}}},\ Vol.~\bibinfo {volume} {2}\ (\bibinfo  {publisher}
  {Courier Corporation},\ \bibinfo {year} {2001})\BibitemShut {NoStop}%
\bibitem [{\citenamefont {Anderson}\ and\ \citenamefont
  {Bergmann}(1951)}]{anderson1951constraints}%
  \BibitemOpen
  \bibfield  {author} {\bibinfo {author} {\bibfnamefont {J.~L.}\ \bibnamefont
  {Anderson}}\ and\ \bibinfo {author} {\bibfnamefont {P.~G.}\ \bibnamefont
  {Bergmann}},\ }\href@noop {} {\bibfield  {journal} {\bibinfo  {journal}
  {Physical Review}\ }\textbf {\bibinfo {volume} {83}},\ \bibinfo {pages}
  {1018} (\bibinfo {year} {1951})}\BibitemShut {NoStop}%
\bibitem [{\citenamefont {Li}\ \emph {et~al.}(2011)\citenamefont {Li},
  \citenamefont {Miao},\ and\ \citenamefont {Miao}}]{Li:2011rn}%
  \BibitemOpen
  \bibfield  {author} {\bibinfo {author} {\bibfnamefont {M.}~\bibnamefont
  {Li}}, \bibinfo {author} {\bibfnamefont {R.-X.}\ \bibnamefont {Miao}},\ and\
  \bibinfo {author} {\bibfnamefont {Y.-G.}\ \bibnamefont {Miao}},\ }\href
  {https://doi.org/10.1007/JHEP07(2011)108} {\bibfield  {journal} {\bibinfo
  {journal} {JHEP}\ }\textbf {\bibinfo {volume} {07}},\ \bibinfo {pages}
  {108}},\ \Eprint {https://arxiv.org/abs/1105.5934} {arXiv:1105.5934 [hep-th]}
  \BibitemShut {NoStop}%
\bibitem [{\citenamefont {Ferraro}\ and\ \citenamefont
  {Guzm{\'a}n}(2018)}]{Ferraro:2018tpu}%
  \BibitemOpen
  \bibfield  {author} {\bibinfo {author} {\bibfnamefont {R.}~\bibnamefont
  {Ferraro}}\ and\ \bibinfo {author} {\bibfnamefont {M.~J.}\ \bibnamefont
  {Guzm{\'a}n}},\ }\href {https://doi.org/10.1103/PhysRevD.97.104028}
  {\bibfield  {journal} {\bibinfo  {journal} {Phys. Rev. D}\ }\textbf {\bibinfo
  {volume} {97}},\ \bibinfo {pages} {104028} (\bibinfo {year} {2018})},\
  \Eprint {https://arxiv.org/abs/1802.02130} {arXiv:1802.02130 [gr-qc]}
  \BibitemShut {NoStop}%
\bibitem [{\citenamefont {Blagojevi{\'c}}\ and\ \citenamefont
  {Nester}(2020)}]{Blagojevic:2020dyq}%
  \BibitemOpen
  \bibfield  {author} {\bibinfo {author} {\bibfnamefont {M.}~\bibnamefont
  {Blagojevi{\'c}}}\ and\ \bibinfo {author} {\bibfnamefont {J.~M.}\
  \bibnamefont {Nester}},\ }\href {https://doi.org/10.1103/PhysRevD.102.064025}
  {\bibfield  {journal} {\bibinfo  {journal} {Phys. Rev. D}\ }\textbf {\bibinfo
  {volume} {102}},\ \bibinfo {pages} {064025} (\bibinfo {year} {2020})},\
  \Eprint {https://arxiv.org/abs/2006.15303} {arXiv:2006.15303 [gr-qc]}
  \BibitemShut {NoStop}%
\bibitem [{\citenamefont {Blixt}\ \emph {et~al.}(2021)\citenamefont {Blixt},
  \citenamefont {Guzm{\'a}n}, \citenamefont {Hohmann},\ and\ \citenamefont
  {Pfeifer}}]{Blixt:2020ekl}%
  \BibitemOpen
  \bibfield  {author} {\bibinfo {author} {\bibfnamefont {D.}~\bibnamefont
  {Blixt}}, \bibinfo {author} {\bibfnamefont {M.-J.}\ \bibnamefont
  {Guzm{\'a}n}}, \bibinfo {author} {\bibfnamefont {M.}~\bibnamefont
  {Hohmann}},\ and\ \bibinfo {author} {\bibfnamefont {C.}~\bibnamefont
  {Pfeifer}},\ }\href {https://doi.org/10.1142/S0219887821300051} {\bibfield
  {journal} {\bibinfo  {journal} {Int. J. Geom. Meth. Mod. Phys.}\ }\textbf
  {\bibinfo {volume} {18}},\ \bibinfo {pages} {2130005} (\bibinfo {year}
  {2021})},\ \Eprint {https://arxiv.org/abs/2012.09180} {arXiv:2012.09180
  [gr-qc]} \BibitemShut {NoStop}%
\bibitem [{\citenamefont {D'Ambrosio}\ \emph {et~al.}(2023)\citenamefont
  {D'Ambrosio}, \citenamefont {Heisenberg},\ and\ \citenamefont
  {Zentarra}}]{DAmbrosio:2023asf}%
  \BibitemOpen
  \bibfield  {author} {\bibinfo {author} {\bibfnamefont {F.}~\bibnamefont
  {D'Ambrosio}}, \bibinfo {author} {\bibfnamefont {L.}~\bibnamefont
  {Heisenberg}},\ and\ \bibinfo {author} {\bibfnamefont {S.}~\bibnamefont
  {Zentarra}},\ }\href {https://doi.org/10.1002/prop.202300185} {\bibfield
  {journal} {\bibinfo  {journal} {Fortsch. Phys.}\ }\textbf {\bibinfo {volume}
  {71}},\ \bibinfo {pages} {2300185} (\bibinfo {year} {2023})},\ \Eprint
  {https://arxiv.org/abs/2308.02250} {arXiv:2308.02250 [gr-qc]} \BibitemShut
  {NoStop}%
\bibitem [{\citenamefont {Hu}\ \emph {et~al.}(2022)\citenamefont {Hu},
  \citenamefont {Katsuragawa},\ and\ \citenamefont {Qiu}}]{Hu:2022anq}%
  \BibitemOpen
  \bibfield  {author} {\bibinfo {author} {\bibfnamefont {K.}~\bibnamefont
  {Hu}}, \bibinfo {author} {\bibfnamefont {T.}~\bibnamefont {Katsuragawa}},\
  and\ \bibinfo {author} {\bibfnamefont {T.}~\bibnamefont {Qiu}},\ }\href
  {https://doi.org/10.1103/PhysRevD.106.044025} {\bibfield  {journal} {\bibinfo
   {journal} {Phys. Rev. D}\ }\textbf {\bibinfo {volume} {106}},\ \bibinfo
  {pages} {044025} (\bibinfo {year} {2022})},\ \Eprint
  {https://arxiv.org/abs/2204.12826} {arXiv:2204.12826 [gr-qc]} \BibitemShut
  {NoStop}%
\bibitem [{\citenamefont {Tomonari}\ and\ \citenamefont
  {Bahamonde}(2024)}]{Tomonari:2023wcs}%
  \BibitemOpen
  \bibfield  {author} {\bibinfo {author} {\bibfnamefont {K.}~\bibnamefont
  {Tomonari}}\ and\ \bibinfo {author} {\bibfnamefont {S.}~\bibnamefont
  {Bahamonde}},\ }\href {https://doi.org/10.1140/epjc/s10052-024-12677-x}
  {\bibfield  {journal} {\bibinfo  {journal} {Eur. Phys. J. C}\ }\textbf
  {\bibinfo {volume} {84}},\ \bibinfo {pages} {349} (\bibinfo {year} {2024})},\
  \bibinfo {note} {[Erratum: Eur.Phys.J.C 84, 508 (2024)]},\ \Eprint
  {https://arxiv.org/abs/2308.06469} {arXiv:2308.06469 [gr-qc]} \BibitemShut
  {NoStop}%
\bibitem [{\citenamefont {Heisenberg}(2025)}]{Heisenberg:2025fxc}%
  \BibitemOpen
  \bibfield  {author} {\bibinfo {author} {\bibfnamefont {L.}~\bibnamefont
  {Heisenberg}},\ }\href@noop {} {\  (\bibinfo {year} {2025})},\ \Eprint
  {https://arxiv.org/abs/2509.18192} {arXiv:2509.18192 [math-ph]} \BibitemShut
  {NoStop}%
\bibitem [{\citenamefont {Einstein}(1955)}]{Einstein1955}%
  \BibitemOpen
  \bibfield  {author} {\bibinfo {author} {\bibfnamefont {A.}~\bibnamefont
  {Einstein}},\ }\href@noop {} {\emph {\bibinfo {title} {The Meaning of
  Relativity}}}\ (\bibinfo  {publisher} {Princeton University Press},\ \bibinfo
  {year} {1955})\BibitemShut {NoStop}%
\bibitem [{\citenamefont {Élie Cartan}(1930)}]{Cartan1930}%
  \BibitemOpen
  \bibfield  {author} {\bibinfo {author} {\bibnamefont {Élie Cartan}},\
  }\href@noop {} {\emph {\bibinfo {title} {La théorie des groupes finis et
  continus et l'analyse situs}}},\ \bibinfo {series} {Mémorial des Sciences
  Mathématiques}, Vol.~\bibinfo {volume} {42}\ (\bibinfo {year}
  {1930})\BibitemShut {NoStop}%
\bibitem [{\citenamefont {Élie Cartan}(1945)}]{Cartan1945}%
  \BibitemOpen
  \bibfield  {author} {\bibinfo {author} {\bibnamefont {Élie Cartan}},\
  }\href@noop {} {\emph {\bibinfo {title} {Les Systèmes Différentielles
  Extérieurs et leurs Applications Géométriques}}}\ (\bibinfo  {publisher}
  {Hermann},\ \bibinfo {address} {Paris},\ \bibinfo {year} {1945})\BibitemShut
  {NoStop}%
\bibitem [{\citenamefont {Kuranishi}(1957)}]{Kuranishi1957}%
  \BibitemOpen
  \bibfield  {author} {\bibinfo {author} {\bibfnamefont {M.}~\bibnamefont
  {Kuranishi}},\ }\href@noop {} {\bibfield  {journal} {\bibinfo  {journal}
  {American Journal of Mathematics}\ }\textbf {\bibinfo {volume} {79}},\
  \bibinfo {pages} {1} (\bibinfo {year} {1957})}\BibitemShut {NoStop}%
\bibitem [{\citenamefont {Seiler}\ and\ \citenamefont
  {Tucker}(1995)}]{Seiler:1995ne}%
  \BibitemOpen
  \bibfield  {author} {\bibinfo {author} {\bibfnamefont {W.~M.}\ \bibnamefont
  {Seiler}}\ and\ \bibinfo {author} {\bibfnamefont {R.~W.}\ \bibnamefont
  {Tucker}},\ }\href {https://doi.org/10.1088/0305-4470/28/15/022} {\bibfield
  {journal} {\bibinfo  {journal} {J. Phys. A}\ }\textbf {\bibinfo {volume}
  {28}},\ \bibinfo {pages} {4431} (\bibinfo {year} {1995})},\ \Eprint
  {https://arxiv.org/abs/hep-th/9506017} {arXiv:hep-th/9506017} \BibitemShut
  {NoStop}%
\bibitem [{\citenamefont {Seiler}(1995)}]{Seiler:1995Fa}%
  \BibitemOpen
  \bibfield  {author} {\bibinfo {author} {\bibfnamefont {W.~M.}\ \bibnamefont
  {Seiler}},\ }\href {https://doi.org/10.1088/0305-4470/28/24/026} {\bibfield
  {journal} {\bibinfo  {journal} {J. Phys. A}\ }\textbf {\bibinfo {volume}
  {28}},\ \bibinfo {pages} {7315} (\bibinfo {year} {1995})}\BibitemShut
  {NoStop}%
\bibitem [{\citenamefont {Seiler}(2000)}]{Seiler2000}%
  \BibitemOpen
  \bibfield  {author} {\bibinfo {author} {\bibfnamefont {W.~M.}\ \bibnamefont
  {Seiler}},\ }\href@noop {} {\bibfield  {journal} {\bibinfo  {journal}
  {Technische Mechanik}\ }\textbf {\bibinfo {volume} {20}},\ \bibinfo {pages}
  {137} (\bibinfo {year} {2000})}\BibitemShut {NoStop}%
\bibitem [{\citenamefont {Seiler}(2010)}]{Seiler2010}%
  \BibitemOpen
  \bibfield  {author} {\bibinfo {author} {\bibfnamefont {W.~M.}\ \bibnamefont
  {Seiler}},\ }\href@noop {} {\emph {\bibinfo {title} {Involution -- The Formal
  Theory of Differential Equations and its Applications in Computer
  Algebra}}},\ \bibinfo {series} {Algorithms and Computation in Mathematics},
  Vol.~\bibinfo {volume} {24}\ (\bibinfo  {publisher} {Springer Berlin
  Heidelberg},\ \bibinfo {year} {2010})\BibitemShut {NoStop}%
\bibitem [{\citenamefont {Adak}\ \emph {et~al.}(2013)\citenamefont {Adak},
  \citenamefont {Sert}, \citenamefont {Kalay},\ and\ \citenamefont
  {Sari}}]{Adak:2008gd}%
  \BibitemOpen
  \bibfield  {author} {\bibinfo {author} {\bibfnamefont {M.}~\bibnamefont
  {Adak}}, \bibinfo {author} {\bibfnamefont {{\"O}.}~\bibnamefont {Sert}},
  \bibinfo {author} {\bibfnamefont {M.}~\bibnamefont {Kalay}},\ and\ \bibinfo
  {author} {\bibfnamefont {M.}~\bibnamefont {Sari}},\ }\href
  {https://doi.org/10.1142/S0217751X13501674} {\bibfield  {journal} {\bibinfo
  {journal} {Int. J. Mod. Phys. A}\ }\textbf {\bibinfo {volume} {28}},\
  \bibinfo {pages} {1350167} (\bibinfo {year} {2013})},\ \Eprint
  {https://arxiv.org/abs/0810.2388} {arXiv:0810.2388 [gr-qc]} \BibitemShut
  {NoStop}%
\bibitem [{\citenamefont {Beltr{\'a}n~Jim{\'e}nez}\ and\ \citenamefont
  {Koivisto}(2022)}]{BeltranJimenez:2022azb}%
  \BibitemOpen
  \bibfield  {author} {\bibinfo {author} {\bibfnamefont {J.}~\bibnamefont
  {Beltr{\'a}n~Jim{\'e}nez}}\ and\ \bibinfo {author} {\bibfnamefont {T.~S.}\
  \bibnamefont {Koivisto}},\ }\href {https://doi.org/10.1142/S0219887822501080}
  {\bibfield  {journal} {\bibinfo  {journal} {Int. J. Geom. Meth. Mod. Phys.}\
  }\textbf {\bibinfo {volume} {19}},\ \bibinfo {pages} {2250108} (\bibinfo
  {year} {2022})},\ \Eprint {https://arxiv.org/abs/2202.01701}
  {arXiv:2202.01701 [gr-qc]} \BibitemShut {NoStop}%
\bibitem [{\citenamefont {Hohmann}(2021)}]{Hohmann:2021ast}%
  \BibitemOpen
  \bibfield  {author} {\bibinfo {author} {\bibfnamefont {M.}~\bibnamefont
  {Hohmann}},\ }\href {https://doi.org/10.1103/PhysRevD.104.124077} {\bibfield
  {journal} {\bibinfo  {journal} {Phys. Rev. D}\ }\textbf {\bibinfo {volume}
  {104}},\ \bibinfo {pages} {124077} (\bibinfo {year} {2021})},\ \Eprint
  {https://arxiv.org/abs/2109.01525} {arXiv:2109.01525 [gr-qc]} \BibitemShut
  {NoStop}%
\bibitem [{\citenamefont {D'Ambrosio}\ \emph {et~al.}(2022)\citenamefont
  {D'Ambrosio}, \citenamefont {Heisenberg},\ and\ \citenamefont
  {Kuhn}}]{DAmbrosio:2021pnd}%
  \BibitemOpen
  \bibfield  {author} {\bibinfo {author} {\bibfnamefont {F.}~\bibnamefont
  {D'Ambrosio}}, \bibinfo {author} {\bibfnamefont {L.}~\bibnamefont
  {Heisenberg}},\ and\ \bibinfo {author} {\bibfnamefont {S.}~\bibnamefont
  {Kuhn}},\ }\href {https://doi.org/10.1088/1361-6382/ac3f99} {\bibfield
  {journal} {\bibinfo  {journal} {Class. Quant. Grav.}\ }\textbf {\bibinfo
  {volume} {39}},\ \bibinfo {pages} {025013} (\bibinfo {year} {2022})},\
  \Eprint {https://arxiv.org/abs/2109.04209} {arXiv:2109.04209 [gr-qc]}
  \BibitemShut {NoStop}%
\bibitem [{\citenamefont {Jensko}(2025)}]{Jensko:2024bee}%
  \BibitemOpen
  \bibfield  {author} {\bibinfo {author} {\bibfnamefont {E.}~\bibnamefont
  {Jensko}},\ }\href {https://doi.org/10.1088/1361-6382/adadbf} {\bibfield
  {journal} {\bibinfo  {journal} {Class. Quant. Grav.}\ }\textbf {\bibinfo
  {volume} {42}},\ \bibinfo {pages} {055011} (\bibinfo {year} {2025})},\
  \Eprint {https://arxiv.org/abs/2407.17568} {arXiv:2407.17568 [gr-qc]}
  \BibitemShut {NoStop}%
\bibitem [{\citenamefont {Heisenberg}\ and\ \citenamefont
  {Hohmann}(2024)}]{Heisenberg:2023tho}%
  \BibitemOpen
  \bibfield  {author} {\bibinfo {author} {\bibfnamefont {L.}~\bibnamefont
  {Heisenberg}}\ and\ \bibinfo {author} {\bibfnamefont {M.}~\bibnamefont
  {Hohmann}},\ }\href {https://doi.org/10.1140/epjc/s10052-024-12810-w}
  {\bibfield  {journal} {\bibinfo  {journal} {Eur. Phys. J. C}\ }\textbf
  {\bibinfo {volume} {84}},\ \bibinfo {pages} {462} (\bibinfo {year} {2024})},\
  \Eprint {https://arxiv.org/abs/2311.05597} {arXiv:2311.05597 [gr-qc]}
  \BibitemShut {NoStop}%
\bibitem [{\citenamefont {Heisenberg}\ \emph {et~al.}(2024)\citenamefont
  {Heisenberg}, \citenamefont {Hohmann},\ and\ \citenamefont
  {Kuhn}}]{Heisenberg:2023wgk}%
  \BibitemOpen
  \bibfield  {author} {\bibinfo {author} {\bibfnamefont {L.}~\bibnamefont
  {Heisenberg}}, \bibinfo {author} {\bibfnamefont {M.}~\bibnamefont
  {Hohmann}},\ and\ \bibinfo {author} {\bibfnamefont {S.}~\bibnamefont
  {Kuhn}},\ }\href {https://doi.org/10.1088/1475-7516/2024/03/063} {\bibfield
  {journal} {\bibinfo  {journal} {JCAP}\ }\textbf {\bibinfo {volume} {03}},\
  \bibinfo {pages} {063}},\ \Eprint {https://arxiv.org/abs/2311.05495}
  {arXiv:2311.05495 [gr-qc]} \BibitemShut {NoStop}%
\bibitem [{\citenamefont {Lue}\ \emph {et~al.}(1999)\citenamefont {Lue},
  \citenamefont {Wang},\ and\ \citenamefont {Kamionkowski}}]{Lue:1998mq}%
  \BibitemOpen
  \bibfield  {author} {\bibinfo {author} {\bibfnamefont {A.}~\bibnamefont
  {Lue}}, \bibinfo {author} {\bibfnamefont {L.-M.}\ \bibnamefont {Wang}},\ and\
  \bibinfo {author} {\bibfnamefont {M.}~\bibnamefont {Kamionkowski}},\ }\href
  {https://doi.org/10.1103/PhysRevLett.83.1506} {\bibfield  {journal} {\bibinfo
   {journal} {Phys. Rev. Lett.}\ }\textbf {\bibinfo {volume} {83}},\ \bibinfo
  {pages} {1506} (\bibinfo {year} {1999})},\ \Eprint
  {https://arxiv.org/abs/astro-ph/9812088} {arXiv:astro-ph/9812088}
  \BibitemShut {NoStop}%
\bibitem [{\citenamefont {Jackiw}\ and\ \citenamefont
  {Pi}(2003)}]{PhysRevD.68.104012}%
  \BibitemOpen
  \bibfield  {author} {\bibinfo {author} {\bibfnamefont {R.}~\bibnamefont
  {Jackiw}}\ and\ \bibinfo {author} {\bibfnamefont {S.-Y.}\ \bibnamefont
  {Pi}},\ }\href {https://doi.org/10.1103/PhysRevD.68.104012} {\bibfield
  {journal} {\bibinfo  {journal} {Phys. Rev. D}\ }\textbf {\bibinfo {volume}
  {68}},\ \bibinfo {pages} {104012} (\bibinfo {year} {2003})}\BibitemShut
  {NoStop}%
\bibitem [{\citenamefont {Alexander}\ and\ \citenamefont
  {Yunes}(2009)}]{Alexander:2009tp}%
  \BibitemOpen
  \bibfield  {author} {\bibinfo {author} {\bibfnamefont {S.}~\bibnamefont
  {Alexander}}\ and\ \bibinfo {author} {\bibfnamefont {N.}~\bibnamefont
  {Yunes}},\ }\href {https://doi.org/10.1016/j.physrep.2009.07.002} {\bibfield
  {journal} {\bibinfo  {journal} {Phys. Rept.}\ }\textbf {\bibinfo {volume}
  {480}},\ \bibinfo {pages} {1} (\bibinfo {year} {2009})},\ \Eprint
  {https://arxiv.org/abs/0907.2562} {arXiv:0907.2562 [hep-th]} \BibitemShut
  {NoStop}%
\bibitem [{\citenamefont {Jenks}\ \emph {et~al.}(2023)\citenamefont {Jenks},
  \citenamefont {Choi}, \citenamefont {Lagos},\ and\ \citenamefont
  {Yunes}}]{Jenks:2023pmk}%
  \BibitemOpen
  \bibfield  {author} {\bibinfo {author} {\bibfnamefont {L.}~\bibnamefont
  {Jenks}}, \bibinfo {author} {\bibfnamefont {L.}~\bibnamefont {Choi}},
  \bibinfo {author} {\bibfnamefont {M.}~\bibnamefont {Lagos}},\ and\ \bibinfo
  {author} {\bibfnamefont {N.}~\bibnamefont {Yunes}},\ }\href
  {https://doi.org/10.1103/PhysRevD.108.044023} {\bibfield  {journal} {\bibinfo
   {journal} {Phys. Rev. D}\ }\textbf {\bibinfo {volume} {108}},\ \bibinfo
  {pages} {044023} (\bibinfo {year} {2023})},\ \Eprint
  {https://arxiv.org/abs/2305.10478} {arXiv:2305.10478 [gr-qc]} \BibitemShut
  {NoStop}%
\bibitem [{\citenamefont {Álvarez}\ \emph {et~al.}(2006)\citenamefont
  {Álvarez}, \citenamefont {Blas}, \citenamefont {Garriga},\ and\
  \citenamefont {Verdaguer}}]{ALVAREZ2006148}%
  \BibitemOpen
  \bibfield  {author} {\bibinfo {author} {\bibfnamefont {E.}~\bibnamefont
  {Álvarez}}, \bibinfo {author} {\bibfnamefont {D.}~\bibnamefont {Blas}},
  \bibinfo {author} {\bibfnamefont {J.}~\bibnamefont {Garriga}},\ and\ \bibinfo
  {author} {\bibfnamefont {E.}~\bibnamefont {Verdaguer}},\ }\href
  {https://doi.org/https://doi.org/10.1016/j.nuclphysb.2006.08.003} {\bibfield
  {journal} {\bibinfo  {journal} {Nuclear Physics B}\ }\textbf {\bibinfo
  {volume} {756}},\ \bibinfo {pages} {148} (\bibinfo {year}
  {2006})}\BibitemShut {NoStop}%
\bibitem [{\citenamefont {Gomes}\ \emph {et~al.}(2024)\citenamefont {Gomes},
  \citenamefont {Beltr{\'a}n~Jim{\'e}nez}, \citenamefont {Cano},\ and\
  \citenamefont {Koivisto}}]{Gomes:2023tur}%
  \BibitemOpen
  \bibfield  {author} {\bibinfo {author} {\bibfnamefont {D.~A.}\ \bibnamefont
  {Gomes}}, \bibinfo {author} {\bibfnamefont {J.}~\bibnamefont
  {Beltr{\'a}n~Jim{\'e}nez}}, \bibinfo {author} {\bibfnamefont {A.~J.}\
  \bibnamefont {Cano}},\ and\ \bibinfo {author} {\bibfnamefont {T.~S.}\
  \bibnamefont {Koivisto}},\ }\href
  {https://doi.org/10.1103/PhysRevLett.132.141401} {\bibfield  {journal}
  {\bibinfo  {journal} {Phys. Rev. Lett.}\ }\textbf {\bibinfo {volume} {132}},\
  \bibinfo {pages} {141401} (\bibinfo {year} {2024})},\ \Eprint
  {https://arxiv.org/abs/2311.04201} {arXiv:2311.04201 [gr-qc]} \BibitemShut
  {NoStop}%
\bibitem [{\citenamefont {D'Ambrosio}\ \emph {et~al.}(2020)\citenamefont
  {D'Ambrosio}, \citenamefont {Garg}, \citenamefont {Heisenberg},\ and\
  \citenamefont {Zentarra}}]{DAmbrosio:2020nqu}%
  \BibitemOpen
  \bibfield  {author} {\bibinfo {author} {\bibfnamefont {F.}~\bibnamefont
  {D'Ambrosio}}, \bibinfo {author} {\bibfnamefont {M.}~\bibnamefont {Garg}},
  \bibinfo {author} {\bibfnamefont {L.}~\bibnamefont {Heisenberg}},\ and\
  \bibinfo {author} {\bibfnamefont {S.}~\bibnamefont {Zentarra}},\ }\href@noop
  {} {\  (\bibinfo {year} {2020})},\ \Eprint {https://arxiv.org/abs/2007.03261}
  {arXiv:2007.03261 [gr-qc]} \BibitemShut {NoStop}%
\end{thebibliography}%

\end{document}